\documentclass[12pt]{article}

\usepackage{graphicx,psfrag,epsf}
\usepackage{enumerate}
\usepackage{natbib}
\usepackage{mathtools}
\usepackage{booktabs}
\usepackage[english]{babel} 
\usepackage{microtype} 
\usepackage{amsmath,amsfonts,amsthm}
\usepackage{chngcntr}
\usepackage[toc,page]{appendix}
\usepackage{ amssymb }
\usepackage{array}
\usepackage{booktabs} 
\usepackage{natbib}
\usepackage{setspace}
\usepackage{soul}
\usepackage{threeparttable}
\usepackage{rotating}
\usepackage{tablefootnote}
\usepackage{color}
\usepackage{tabularx}
\usepackage[font = small,labelfont=bf,textfont=it]{caption} 
\usepackage{footnote}
\usepackage{algorithm}
\usepackage[noend]{algpseudocode}
\usepackage{footnote}
\usepackage{multirow}

\makeatletter
\newcommand{\distas}[1]{\mathbin{\overset{#1}{\kern\z@\sim}}}%
\newcommand{\bm}[1]{\mathbf{#1}}
\newsavebox{\mybox}\newsavebox{\mysim}
\newcommand{\distras}[1]{%
  \savebox{\mybox}{\hbox{\kern3pt$\scriptstyle#1$\kern3pt}}%
  \savebox{\mysim}{\hbox{$\sim$}}%
  \mathbin{\overset{#1}{\kern\z@\resizebox{\wd\mybox}{\ht\mysim}{$\sim$}}}%
}
\newtheorem{theorem}{Theorem}

\newtheorem{definition}{Definition}

\newtheorem{lemma}[theorem]{Lemma}
\newtheorem{corollary}{Corollary}
\newtheorem{proposition}{Proposition}

\newcolumntype{C}[1]{>{\centering\let\newline\\\arraybackslash\hspace{0pt}}m{#1}}

\newcommand{\be}{\begin{equation}}
\newcommand{\ee}{\end{equation}}
\newcommand{\bi}{\begin{itemize}}
\newcommand{\ei}{\end{itemize}}
\newcommand{\ben}{\begin{enumerate}}
\newcommand{\een}{\end{enumerate}}
\newcommand{\stb}{\State $\bullet$ \;}

\allowdisplaybreaks
\DeclareMathOperator*{\argmin}{\arg\!\min}
\makeatother

\let\oldbibliography\thebibliography
\renewcommand{\thebibliography}[1]{\oldbibliography{#1}
\setlength{\itemsep}{0pt}} 

\newcommand{\blind}{1}

\patchcmd{\footnotemark}{\stepcounter{footnote}}{\refstepcounter{footnote}}{}{}
\newcounter{savecntr}
\newcounter{restorecntr}

\begin{document}

\def\spacingset#1{\renewcommand{\baselinestretch}%
{#1}\small\normalsize} \spacingset{1}

\if1\blind
{
  \title{\bf \texttt{cmenet}: a new method for bi-level variable selection of conditional main effects}
  \small
  \author{Simon Mak\setcounter{savecntr}{\value{footnote}}\thanks{School of Industrial and Systems Engineering, Georgia Institute of Technology}, \; C. F. Jeff Wu\setcounter{restorecntr}{\value{footnote}}%
  \setcounter{footnote}{\value{savecntr}}\footnotemark
  \setcounter{footnote}{\value{restorecntr}}
\footnote{Corresponding author} 
}
  \maketitle
} \fi

\if0\blind
{
  \title{\bf \texttt{cmenet}: a new method for bi-level variable selection of conditional main effects}
   \date{}
   \maketitle
} \fi

\bigskip

\begin{abstract}
This paper introduces a novel method for selecting main effects and a set of reparametrized effects called conditional main effects (CMEs), which capture the conditional effect of a factor at a fixed level of another factor. CMEs represent interpretable, domain-specific phenomena for a wide range of applications in engineering, social sciences and genomics. The key challenge is in incorporating the implicit grouped structure of CMEs within the variable selection procedure itself. We propose a new method, \texttt{cmenet}, which employs two principles called CME coupling and CME reduction to effectively navigate the selection algorithm. Simulation studies demonstrate the improved CME selection performance of \texttt{cmenet} over more generic selection methods. Applied to a gene association study on fly wing shape, \texttt{cmenet} not only yields more parsimonious models and improved predictive performance over standard two-factor interaction analysis methods, but also reveals important insights on gene activation behavior, which can be used to guide further experiments. Efficient implementations of our algorithms are available in the R package \textsc{cmenet} in CRAN.
\end{abstract}

\noindent%
\vfill

\newpage
\spacingset{1.45} 

\section{Introduction}
This paper proposes a new method for selecting \textit{main effects} (MEs) and a set of reparametrized effects called \textit{conditional main effects} (CMEs) from observational data. A CME can be described as follows. Let $A$ and $B$ denote two binary factors with levels $+$ and $-$. The CME $A|B+$ is then defined as the effect $A$ when effect $B$ is at the $+$ level, and 0 when $B$ is at the $-$ level. In words, such an effect quantifies the influence of $A$ \textit{only} when $B$ is at the level $+$. The CME $A|B-$ can be defined analogously.


The appeal for CMEs as basis functions for variable selection comes from its interpretability in a wide range of applications, including genomics and the social sciences. For example, in gene association studies, where the goal is to identify important genetic contributions for a trait or disease, the CME $A|B+$ quantifies the significance of gene $A$ \textit{only when} gene $B$ is present. Such conditional effects are biologically interpretable and meaningful, as noted in \cite{CD2013}: ``[the examination] of how one mutation behaves when in the presence of a second mutation forms the basis of our understanding of genetic interactions, and is part of the fundamental toolbox of genetic analysis.'' Viewed this way, the selection of CMEs can therefore serve as an effective tool for investigating the activation and inhibition behavior of gene-gene interactions, namely, which genes are \textit{conditionally} active, and which are important in \textit{activating} or \textit{inhibiting} other genes. CMEs also arise naturally in many engineering applications. For example, in an injection molding experiment with two settings for mold temperature $A$ and holding pressure $B$ (pg. 352 of \citealp{Mon2008}), the CME $A|B+$ measures the effectiveness of mold temperature \textit{only at} a high level of holding pressure. This conditional effect may be a result of material properties for the molding liquid, and the discovery of such effects can provide valuable insight on the injection process.

The idea of CMEs was first introduced in \cite{Wu2015} as a way to disentangle effects which are fully-aliased (i.e., perfectly correlated) in a \textit{designed} experiment. Ever since the pioneering work of \cite{Fin1945}, it has been widely accepted in the design community that fully-aliased effects in a regular, two-level design cannot be ``de-aliased'' without adding more experimental runs. Such a belief was shown to be false in \cite{Wu2015}, where the author employed a reparametrization of these fully-aliased effects into CMEs, and allowed for the selection of the resulting conditional effects. A variable selection method for designed experiments is further developed in \cite{SW2016}, making use of the natural groupings of CMEs into so-called twin, sibling and family effects. In this paper, we generalize this CME selection framework to \textit{observational data}, by exploiting the implicit structure of CMEs to form new effect groups and to motivate a novel penalized selection criterion.

For penalized variable selection methods, the usual procedure for two-level factors is to first normalize each factor to zero mean and unit variance \citep{Tib1997}. Treating these rescaled factors as continuous variables, standard variable selection techniques using the $l_1$-penalty in LASSO \citep{Tib1997} or non-convex penalties (e.g., \citealp{FF1993, FL2001, Zha2010}) can then be used to identify significant effects. For the problem at hand, however, such methods are inappropriate, because they do not account for the implicit group structure present in CMEs. Grouped selection techniques, such as the group LASSO \citep{YL2006} or the overlapping group LASSO \citep{Jea2009}, are also not suitable here, because such methods select \textit{all} effects from an active group, whereas only a handful of effects may be active within a CME group.

In this light, a \textit{bi-level} selection strategy is needed to select both \textit{active CME groups} and \textit{active effects within CME groups}. In recent years, there have been important developments on bi-level variable selection, including the sparse group LASSO \citep{WL2008, Sea2013} and the group exponential LASSO \citep{BH2009, Bre2015}. We extend the latter framework here, because it allows us to encode within the penalization criterion two important selection principles called \textit{CME coupling} and \textit{CME reduction}. These two principles guide the search for good CME models, and can be seen as an extension of effect heredity and effect hierarchy \citep{WH2011}, two guiding principles used for model selection in designed experiments.

The paper is organized as follows. Section 2 provides some motivation for the problem at hand, including the implicit collinearity structure of CME groups and its effect on selection inconsistency. Section 3 proposes a new penalization criterion for CME selection, and illustrates two appealing selection principles (CME coupling and CME reduction) encoded within this criterion. Section 4 introduces a coordinate descent optimization algorithm using threshold operators, and presents an efficient tuning procedure for penalty parameters. Section 5 outlines several simulations comparing the CME selection performance of \texttt{cmenet} to existing variable selection methods. Section 6 then demonstrates the usefulness of the proposed method in a gene association study, and Section 7 concludes with directions for future research.

\section{Background and motivation}
\label{sec:bm}

\subsection{CME and CME groups}
\label{sec:cmegroups}

We first define some notation. Let $\bm{y} \in \mathbb{R}^n$ be a vector of $n$ observations, and suppose $p$ main effects are considered. {For effect $J$, let $\tilde{\bm{x}}_j = (x_{1,j}, \cdots, x_{n,j})$ $\in \{-1,+1\}^n$ be its binary covariate vector, $j = 1, \cdots, p$.} The tilde on $\tilde{\bm{x}}_j$ distinguishes the binary covariate from its normalized analogue $\bm{x}_j$, which is introduced later. A CME can then be defined as follows:
\begin{definition}[Conditional main effect]
The \textup{conditional main effect (CME)} of $J$ given $K$ at level +, denoted as $J|K+$, quantifies the effect of covariate vector $\tilde{\bm{x}}_{j|k+} = (\tilde{{x}}_{1,j|k+}, \cdots, \tilde{{x}}_{n,j|k+})$, where:
\[\tilde{{x}}_{i,j|k+} = \begin{dcases}
\tilde{x}_{i,j}, & \text{ if $\tilde{x}_{i,k} = +1$}\\
0, & \text{ if $\tilde{x}_{i,k} = -1$}
\end{dcases}, \quad \text{ for } i = 1, \cdots, n.\]
The CME $J|K-$ can be defined in a similar manner.
\label{def:cme}
\end{definition}
\noindent Throughout this paper, the effects $J$ and $K$ {are respectively referred to as} the \textit{parent effect} and the \textit{conditioned effect} of $J|K+$. Using this terminology, $J|K+$ quantifies the effect of parent $J$, given its conditioned effect $K$ is at level +. For illustration, Table \ref{tbl:cme} shows the four possible CMEs constructed from two main effects $A$ and $B$.

\begin{table}[t]
\small
\centering
\begin{tabular}{c | c || c | c | c | c}
\toprule
$A$ & $B$ & $A|B+$ & $A|B-$ & $B|A+$ & $B|A-$\\
\toprule
+1 & +1 & +1 & 0 & +1 & 0\\
+1 & -1 & 0 & +1 & -1 & 0\\
-1 & +1 & -1 & 0 & 0 & +1\\
-1 & -1 &  0 & -1 & 0 & -1\\
\toprule
\end{tabular}
\caption{{Model matrix for the two MEs $A$ and $B$, and its four CMEs $A|B+,A|B-,B|A+,B|A-$.}}
\label{tbl:cme}
\end{table}

Restricted to two-level, fractional factorial designed experiments, \cite{SW2016} identified three important CME groups for selecting an \textit{orthogonal} model, in which active effects are orthogonal to each other. These three groups are: (a) \textit{sibling} CMEs: CMEs with the same parent effect, (b) \textit{twin} CMEs: CME pairs with the same parent and conditioned effect, but with the sign for the latter flipped, (c) \textit{family} CMEs: CMEs with fully-aliased interaction effects. Leveraging this group structure, \cite{SW2016} proposed three rules for selecting a parsimonious and orthogonal model. Rule 1 (the most important selection rule) relies on the two simple mathematical identities:
\begin{equation}
\tilde{\bm{x}}_{j|k+} = \frac{1}{2}\left(\tilde{\bm{x}}_{j} + \tilde{\bm{x}}_{j * k} \right) \quad \text{and} \quad \tilde{\bm{x}}_{j|k-} = \frac{1}{2}\left(\tilde{\bm{x}}_{j} - \tilde{\bm{x}}_{j * k} \right).
\label{eq:cmeconstr}
\end{equation}
Here, $\tilde{\bm{x}}_{j * k} = \tilde{\bm{x}}_j \circ \tilde{\bm{x}}_k$ is the covariate vector for the traditional two-factor interaction (2FI) $J * K$, where $\circ$ is the Hadamard (entry-wise) product. From \eqref{eq:cmeconstr}, the CME $J|K+$ can then be viewed as an average of the main effect for $J$ and the interaction effect for $J * K$; a similar interpretation holds for the CME $J|K-$. Motivated by this interpretation, Rule 1 of \cite{SW2016} replaces a selected ME $J$ and 2FI $J * K$ with either (a) the CME $J|K+$, if the signs for $J$ and $J * K$ are identical and their effect magnitudes are similar, or (b) the CME $J|K-$, if the signs for $J$ and $J*K$ are different and their effect magnitudes are similar. Such a rule (along with Rules 2 and 3) allows for the disentangling of fully-aliased interaction effects in a designed experiment.

The above CME groupings, however, are not suitable for analyzing observational data, because an orthogonal model is most likely not attainable for this more general setting. Instead, by exploring the correlation structure of CMEs, the following new groupings can be derived:
\ben
\item \textit{Sibling} CMEs: CMEs which share the same parent effect, e.g., $\{A|B+, A|B-, A|C+, $ \par $A|C-, A|D+, A|D-, \cdots\}$. This is the same as in \cite{SW2016}.
\item \textit{Parent-child} pairs: An effect pair consisting of a CME and its parent ME, e.g., $\{A|B+, A\}, \{A|C+, A\}, \cdots$.
\item \textit{Cousin} CMEs\footnote{From a purely linguistic point-of-view, these effects are not cousins, because their parent effects are unrelated. However, the notion of cousin nicely encapsulates a weaker form of a sibling relationship, which is the intended meaning here.}: CMEs which share the same conditioned effect, e.g., $\{B|A+, B|A-,$ \par $C|A+, C|A-, D|A+, D|A-, \cdots\}$.
\een
We first outline the justification for these groups in terms of collinearity, then discuss why such groupings are appealing from a selection consistency perspective.

\subsection{Group structure for collinearity}
\label{sec:group}
To explore the group structure of CMEs, consider the following latent model for the main effects $\{\tilde{\bm{x}}_j\}_{j=1}^p \subseteq \{-1,+1\}^n$. Define the latent matrix $\bm{Z} = {(z_{i,j})_{i=1}^n}_{j=1}^p \in \mathbb{R}^{n \times p}$, where each row of $\bm{Z}$ is drawn independently from the equicorrelated normal distribution $\mathcal{N}\{\bm{0},\rho \bm{J}_p + (1-\rho) \bm{I}_p\}$. Here, $\bm{I}_p$ is the $p \times p$ identity matrix, $\bm{J}_p$ is the $p \times p$ matrix of ones, and $\rho \in [0,1]$. We then assume the following form for the binary covariates $\{\tilde{\bm{x}}_j\}_{j=1}^p$:
\begin{equation}
\tilde{x}_{i,j} = \mathbf{1}\{z_{i,j} > 0\} - \mathbf{1}\{z_{i,j} \leq 0\}, \quad i = 1, \cdots, n, \;  j = 1, \cdots, p.
\label{eq:latent}
\end{equation}
Note that a larger value of $\rho$ induces a higher correlation between the binary main effects.

\begin{figure}[t]
\centering
\includegraphics[width=0.6\textwidth]{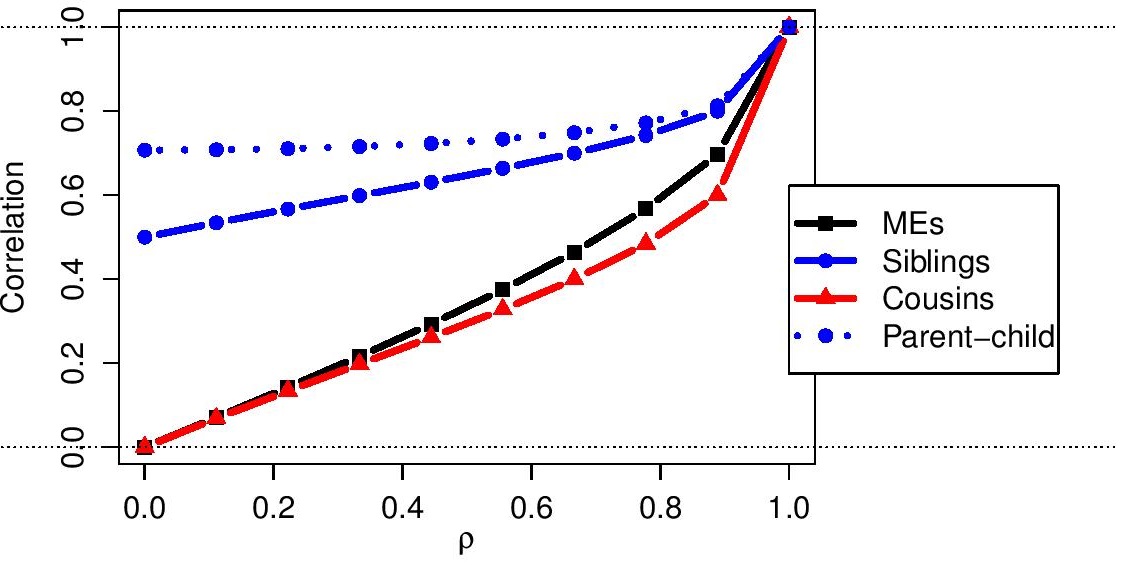}
\caption{Pairwise correlations within the four effect groups as a function of latent correlation $\rho$.}
\label{fig:corr}
\end{figure}

Without loss of generality, assume here that the conditioned effects are set at the + level for all CMEs. With the above model, the following theorem reveals an interesting group structure for CMEs. For brevity, proofs of all technical results are deferred to the Appendix.	
\begin{theorem}[Pairwise correlation within groups]
Under the latent model \eqref{eq:latent} for main effects, the four effect groups have the following pairwise correlations:
\begin{center}
\small
\begin{tabular}{c c || c c}
\toprule
\textit{Group} & \textit{Pairwise correlation} & \textit{Group} & \textit{Pairwise correlation}\\
\toprule
\textup{Main effects} & $\frac{2\sin^{-1} \rho}{\pi}$ & \textup{Parent-child} & $\frac{1}{2\sigma_c}$ \\
\textup{Siblings} & $\frac{1}{\sigma^{2}_c}\left\{ \frac{1}{4} + \frac{\sin^{-1}\rho}{2\pi} - \left(\frac{\sin^{-1}\rho}{\pi}\right)^2 \right\}$ & \textup{Cousins} & $\frac{1}{\sigma^{2}_c} \left\{ \frac{\sin^{-1}\rho}{\pi} - \left(\frac{\sin^{-1}\rho}{\pi}\right)^2 \right\}$ \\
\toprule
\end{tabular}
\end{center}
where $\sigma^2_c = 1/2 - (\sin^{-1} \rho / \pi)^2$.
\label{thm:corr}
\end{theorem}

Figure \ref{fig:corr} plots the pairwise correlations in Theorem \ref{thm:corr} as a function of the latent correlation parameter $\rho$. Two key observations can be made. First, the magnitudes of these correlations impose a natural hierarchy on the effect groups. For all values of $\rho \in (0,1)$, parent-child pairs have the largest correlations, followed by sibling pairs, then main effect and cousin pairs. Second, the correlation group structure can vary considerably for different choices of $\rho$. In the independent setting of $\rho = 0$, sibling and parent-child pairs exhibit high correlations of $0.5$ and $1/\sqrt{2}$ $(\approx 0.71)$, respectively, whereas the remaining two groups have zero correlation. For moderately large choices of $\rho$, say, $\rho = 1/\sqrt{2}$ $(\approx 0.71)$, these correlations become larger and more distinct between different groups, thereby amplifying the underlying CME group structure.

In light of this complex collinearity structure, one may suspect that standard variable selection techniques, such as the LASSO, would perform poorly for CME selection, because such methods impose the same regularization penalty over all variables, and ignore the implicit grouped correlation structure. This is indeed the case, and we demonstrate its poor selection performance in the following section and in the simulations of Section \ref{sec:sim}.

\subsection{Selection inconsistency}
\label{sec:inc}
An important property of a selection method is its \textit{consistency} in choosing the correct model. Put mathematically, a method is (sign-)\textit{selection consistent} if $\lim_{n \rightarrow \infty} \mathbb{P}(\hat{\boldsymbol{\beta}}_n =_s \boldsymbol{\beta}) = 1$, where $\boldsymbol{\beta} \in \mathbb{R}^p$ is the true coefficient vector, $\hat{\boldsymbol{\beta}}_n$ is the estimated vector from $n$ observations, and $=_s$ denotes equality in sign (see \citealp{ZB2006} for a precise definition). The following theorem shows that LASSO is indeed inconsistent for simple CME models:
\begin{theorem}[Selection inconsistency of LASSO]
Under the latent model \eqref{eq:latent}, the LASSO is selection inconsistent in the following situations: (a) for $\rho \geq 0$, a model with $q \geq 3$ active siblings, (b) for $\rho \geq 0.27$, a model with $q=2$ active main effects, and (c) for $\rho \geq 0.29$, a model with $q \geq 6$ active cousins.
\label{thm:inc}
\end{theorem}
\noindent Theorem \ref{thm:inc} demonstrates the poor selection of LASSO for simple CME models, even when little-to-no latent correlation is present. Part (a) says that, even in the uncorrelated setting of $\rho = 0$, LASSO yields poor selection whenever three (or more) siblings are present; part (b) says that, for mild correlations as low as 0.27, the same poor selection arises for two active MEs; part (c) says that, for correlations lower than 0.29, LASSO enjoys good selection even when many cousins (up to 5) are active -- this is not too surprising, because cousins experience the lowest pairwise correlations of the four groups. The proof of this theorem relies on the \textit{irrepresentability condition} \citep{ZB2006}, which shows that the LASSO is selection inconsistent when active variables are highly correlated with non-active ones.

\section{\texttt{cmenet}: Penalization framework}
To address these selection concerns, we propose a novel bi-level variable selection method called \texttt{cmenet}, which can identify both active CME groups and active effects within such groups. Similar to popular selection methods such as the elastic net \citep{ZH2005} and \texttt{SparseNet} \citep{Mea2012}, the name \texttt{cmenet} draws an analogy between the proposed method's ability to select active variables amongst non-active ones, and a fishing net's ability to catch larger fish amongst smaller ones. The penalization scheme for \texttt{cmenet} encodes two important principles, called \textit{CME coupling} and \textit{CME reduction}, which, as we show in this section, help guide the selection procedure for CMEs. 

\subsection{Selection criterion}
\label{sec:crit}
We first introduce the selection criterion. Let $\bm{x}_j \in \mathbb{R}^n$ be the normalized vector for the binary main effect covariate $\tilde{\bm{x}}_j$, with $\bm{x}_j^T \bm{1}_n = 0$ and $n^{-1}\|\bm{x}_{j}\|_2^2 = 1$, along with a similar notation for CME covariates. Further let {$\bm{X} = ({\bm{x}}_1, \cdots, {\bm{x}}_{p'}) \in \mathbb{R}^{n \times p'}$} be the full model matrix consisting of these normalized ME and CME effects, where $p' = p + 4 {p \choose 2}$ is the total {number of effects} considered. For simplicity, assume all considered effects are MEs and CMEs for the following exposition; Section \ref{sec:alg} gives a simple extension for selecting these effects along with \textit{other} covariate factors. Let $\boldsymbol{\beta} \in \mathbb{R}^{p'}$ be the coefficient vector, with $\beta_j$ and $\beta_{j|k+}$ its corresponding coefficients for ME $J$ and CME $J|K+$. Finally, assume that $\bm{y}$ is centered, i.e., $\bm{y}^T\bm{1}_n = 0$. 

For effect groups, define $\mathcal{S}(j) = \{J, J|A+, J|A-, J|B+, J|B-, \cdots\}$ as the \textit{sibling group} for parent effect $j$, and $\mathcal{C}(j) = \{J, A|J+, A|J-, B|J+, B|J-, \cdots\}$ as the \textit{cousin group} for conditioned effect $j$, $j = 1, \cdots, p$. We propose the following selection criterion, which can be viewed as an extension of the hierarchical framework in \cite{BH2009}:
\begin{align}
\begin{split}
\min_{\boldsymbol{\beta}}Q(\boldsymbol{\beta}) &\equiv \min_{\boldsymbol{\beta}} \left\{\frac{1}{2n}\| \bm{y} - \bm{X}\boldsymbol{\beta}\|^2_2 + P_{\mathcal{S}}(\boldsymbol{\beta}) + P_{\mathcal{C}}(\boldsymbol{\beta}) \right\},\\
P_{\mathcal{S}}(\boldsymbol{\beta}) & \equiv \sum_{j=1}^p f_{o,\mathcal{S}}\left\{ \sum_{k \in \mathcal{S}(j)} f_{i,\mathcal{S}}\left( {\beta}_k \right)\right\}, \quad P_{\mathcal{C}}(\boldsymbol{\beta}) \equiv \sum_{j=1}^p f_{o,\mathcal{C}}\left\{ \sum_{k \in \mathcal{C}(j)} f_{i,\mathcal{C}}\left( \beta_k \right) \right\}.
\label{eq:crit}
\end{split}
\end{align}
Here, $f_{o, \mathcal{S}}$ and $f_{i, \mathcal{S}}$ (similarly, $f_{o, \mathcal{C}}$ and $f_{i, \mathcal{C}}$) are \textit{outer} and \textit{inner} penalties which control the \textit{between-group} and \textit{within-group} selection for sibling (similarly, cousin) groups, respectively. While the specific penalty functions are left arbitrary in \eqref{eq:crit}, we will introduce \texttt{cmenet} for the specific choice of the exponential penalty in \cite{Bre2015} for outer penalty, and the (scaled) minimax concave-plus penalty (MC+) in \cite{Zha2010} for inner penalty:
\begin{align}
\begin{split}
&\textit{Outer: }f_{o, \mathcal{S}}(\theta) = \eta_{\lambda_s,\tau}(\theta), \; f_{o, \mathcal{C}}(\theta) = \eta_{\lambda_c,\tau}(\theta), \; \text{where } \eta_{\lambda,\tau}(\theta) = \frac{\lambda^2}{\tau} \left\{ 1 - \exp\left( - \frac{\tau \theta}{\lambda} \right) \right\},\\
&\textit{Inner: } f_{i, \mathcal{S}}(\beta) = g_{\lambda_s,\gamma}(\beta), \; f_{i, \mathcal{C}}(\beta) = g_{\lambda_c,\gamma}(\beta), \; \text{where } g_{\lambda,\gamma}(\beta) = \int_{0}^{|\beta|} \left( 1 - \frac{x}{\lambda \gamma}\right)_+ dx.
\end{split}
\label{eq:expomc}
\end{align}
This inner penalty is a scaled version of the MC+ penalty $\lambda g_{\lambda,\gamma}(\beta)$ in \cite{Zha2010} without the scaling factor $\lambda$; such a factor is accounted for in the outer exponential penalty $\eta_{\lambda,\tau}(\theta)$. 

The appeal for the ``exponential-MC+'' framework in \eqref{eq:expomc} is that it provides a concise parametrization of the grouped collinearity structure in Section \ref{sec:bm}. First, the penalty parameters $\lambda_s > 0$ and $\lambda_c > 0$ allow for differing regularization within sibling and cousin groups, respectively, with larger penalty values reducing the number of selected effects in each group. Assuming such parameters are tuned via cross-validation, a smaller tuned value of $\lambda_s$ suggests many sibling effects are present in the data, while a smaller $\lambda_c$ suggests the same for cousin effects. Second, the parameter $\gamma > 1$ controls the non-convexity of the inner MC+ penalty, and provides a ``bridge'' between the $l_0$-penalty (obtained when $\gamma \rightarrow 1^+$) and the $l_1$-penalty in LASSO (obtained when $\gamma \rightarrow \infty$). In view of the selection problems for LASSO (see Theorem \ref{thm:inc}), such a parameter allows for improved selection of the highly correlated CMEs, say, within a sibling group. Lastly, the parameter $\tau$ provides two appealing principles called CME coupling and reduction, which we introduce below.

\subsection{CME coupling and reduction}
\label{sec:couple}
Consider first a CME $J|K+$ which has yet to be selected, and assume without loss of generality that $\bm{x}_{j|k+}^T(\bm{y}-\bm{X}\boldsymbol{\beta})/n> 0$. Taking the derivative of $Q(\boldsymbol{\beta})$ with respect to $\beta_{j|k+}$, and setting $\beta_{j|k+} = 0$ (as $J|K+$ is not in the model), we get:
\begin{align}
\begin{split}
\frac{\partial}{\partial \beta_{j|k+}} Q(\boldsymbol{\beta})\Big|_{\beta_{j|k+}=0} &= -\frac{1}{n} \bm{x}_{j|k+}^T(\bm{y}-\bm{X}\boldsymbol{\beta}) + \Delta_{\mathcal{S}(j)} + \Delta_{\mathcal{C}(k)},\\
\text{where } \; \Delta_{\mathcal{S}(j)} &= \lambda_s \exp\left\{ -\frac{\tau \|\boldsymbol{\beta}_{\mathcal{S}(j)}\|_{\lambda_s,\gamma}}{\lambda_s} \right\} \text{ and } \Delta_{\mathcal{C}(k)} = \lambda_c \exp\left\{ -\frac{\tau \|\boldsymbol{\beta}_{\mathcal{C}(k)}\|_{\lambda_c,\gamma}}{\lambda_c} \right\}.
\label{eq:couple}
\end{split}
\end{align}
Here, $\boldsymbol{\beta}_g \in \mathbb{R}^{|g|}$ denotes the coefficient vector for {an} effect subset $g \subseteq \{1, \cdots, p'\}$, and $\|\boldsymbol{\beta}_g\|_{\lambda,\gamma} \equiv \sum_{l \in g} g_{\lambda,\gamma}(\beta_l)$ denotes its ``norm'' under the inner MC+ penalty. (For completeness, a full derivation of the subgradient for $Q(\boldsymbol{\beta})$ -- which is quite technical and requires several applications of the chain rule -- is found in equation \eqref{eq:subgrad} of the Appendix.)

Equation \eqref{eq:couple} reveals an appealing selection property of \texttt{cmenet} called CME coupling, which we describe below. Note that, when more effects have been selected in the sibling group $\mathcal{S}(j)$ (or cousin group $\mathcal{C}(k)$), the effect norms $\|\boldsymbol{\beta}_{\mathcal{S}(j)}\|$ (or $\|\boldsymbol{\beta}_{\mathcal{C}(k)}\|$) become larger. This then results in a smaller linearized slope $\Delta_{\mathcal{S}(j)}$ (or $\Delta_{\mathcal{C}(k)}$), which generates a decrease in the derivative $\frac{\partial}{\partial \beta_{j|k+}} Q(\boldsymbol{\beta})$ in \eqref{eq:couple}. Since the goal is to minimize the selection criterion $Q(\boldsymbol{\beta})$, a smaller derivative allows for greater decrease in $Q(\boldsymbol{\beta})$ when $\beta_{j|k+}$ enters the model. In other words, the CME $J|K+$ has a greater chance of entering the model when other effects in its sibling group $\mathcal{S}(j)$ or its cousin group $\mathcal{C}(k)$ have already been selected; the selection of sibling or cousin effects can \textit{couple} in the selection of the CME $J|K+$. We call this property \textit{CME coupling}, following the idea of effect coupling in \cite{Bre2015}.

Consider next a ME $J$ which has yet to be selected, and assume again that $\bm{x}_{j}^T(\bm{y}-\bm{X}\boldsymbol{\beta})/n > 0$. Taking the derivative of $Q(\boldsymbol{\beta})$ with respect to $\beta_j$, and setting $\beta_j = 0$ (as $J$ is not in the model), we get:
\begin{align}
\begin{split}
\frac{\partial}{\partial \beta_j} Q(\boldsymbol{\beta}) \Big|_{\beta_j=0} = -\frac{1}{n} \bm{x}_j^T(\bm{y}-\bm{X}\boldsymbol{\beta}) + \Delta_{\mathcal{S}(j)} + \Delta_{\mathcal{C}(j)}.
\label{eq:reduce}
\end{split}
\end{align}
The interpretation of equation \eqref{eq:reduce} is similar to that for \eqref{eq:couple}. When more effects have already been selected in the sibling group $\mathcal{S}(j)$ (or the cousin group $\mathcal{C}(j)$), the linearized slopes $\Delta_{\mathcal{S}(j)}$ (or $\Delta_{\mathcal{C}(j)}$) become smaller, which then decreases the derivative $\frac{\partial}{\partial \beta_j} Q(\boldsymbol{\beta})$ in \eqref{eq:reduce}. Hence, the ME $J$ enters the model more easily when effects in its sibling group $\mathcal{S}(j)$ or its cousin group $\mathcal{C}(j)$ have already been selected; the selection of many sibling or cousin effects can then \textit{reduce} to its underlying main effect. We refer to this phenomenon as \textit{CME reduction}.

The notions of CME coupling and reduction are quite intuitive to expect in many CME applications. Consider the gene expression example in the Introduction, where the selection of the CME $A|B+$ indicates the effectiveness of gene $A$ only when gene $B$ is present. When several sibling CMEs of $A$, say, $A|B+$ and $A|C+$, are already selected in the model, one naturally expects gene $A$ to be conditionally active under more genes as well. In other words, conditional effects with parent $A$ are more likely to be active compared to conditional effects with no selected siblings -- this is precisely the principle of CME coupling. However, when many sibling effects of gene $A$ have already been selected, one may suspect that the underlying parent effect for gene $A$ is active instead of these selected siblings -- this is precisely the principle of CME reduction. A similar intuition holds for cousin effects.

An interesting parallel can also be made connecting CME coupling and reduction with the two guiding principles for model selection in designed experiments \citep{WH2011}. The first principle, called (weak) \textit{effect heredity}, states that higher-order interactions can be selected only when either of its parent main effects are in the model. This idea is quite similar to CME coupling, which allows for easier selection of a CME when effects with either the same parent or conditioned ME have been selected. Furthermore, note that a CME can be interpreted as a component of an interaction effect, because the difference of the two CMEs $A|B+$ and $A|B-$ is precisely the two-factor interaction $A * B$ \citep{SW2016}. Coupling can therefore be seen as an \textit{extension} of effect heredity, after breaking an interaction effect (which is often difficult to interpret) into more interpretable conditional effects. The second principle, called \textit{effect hierarchy}, states that lower-order interactions are more likely active than higher-order ones. This is akin to CME reduction, which encourages the reduction of selected sibling (or cousin) CMEs to its parent (conditioned) effect when too many siblings (cousins) are in the model.

\section{\texttt{cmenet}: Optimization framework}

With the proposed penalty $Q(\boldsymbol{\beta})$ in hand, we now present an optimization framework for \texttt{cmenet} in three parts. We first introduce the optimization algorithm for minimizing $Q(\boldsymbol{\beta})$, then describe several computational techniques for tuning penalty parameters, and finally conclude with several novel CME screening rules for speeding up the tuning procedure.

\subsection{Optimization algorithm}
\subsubsection{Coordinate descent and threshold operators}
\label{sec:thresh}
We first develop the algorithmic framework for minimizing the selection criterion $Q(\boldsymbol{\beta})$. A key tool in this optimization algorithm is \textit{coordinate descent}, which can be explained as follows. Viewing $Q(\boldsymbol{\beta})$ as a function of only the first coefficient $\beta_1$ (call this $Q_1(\beta_1)$), we first update $\beta_1$ as the minimizer of $Q_1(\cdot)$, keeping the remaining $p'-1$ coefficients fixed. The same procedure is then applied cyclically over $\beta_2, \cdots, \beta_{p'}$, and repeated until the full coefficient vector $\boldsymbol{\beta}$ converges. In recent years, coordinate descent has become widely used in the variable selection literature (see, e.g., \citealp{Fu1998, Fea2007,Mea2011}), due to its simplicity and efficiency for high-dimensional problems. The key to efficiency lies in the existence of a \textit{closed-form} minimizer for the coordinate-wise objective $Q_j(\cdot)$, also known as a \textit{threshold function} from signal processing \citep{Don1995}. We derive below such a threshold function for $Q(\boldsymbol{\beta})$.

Before delving into details, we first investigate the convexity properties of $Q(\boldsymbol{\beta})$:
\begin{proposition}[Strict convexity]
$Q(\boldsymbol{\beta})$ is strictly convex whenever $\tau + 1/\gamma< {\lambda_{min}(\bm{X}^T\bm{X})}/(2n)$, where $\lambda_{min} (\cdot)$ returns the minimum eigenvalue. Also, assuming each column $\bm{x}_j$ of $\bm{X}$ is normalized (i.e., $\bm{x}_j^T\bm{1} = 0$ and $n^{-1}\|\bm{x}_j\|_2^2 = 1$ for any $j = 1, \cdots, p'$), it follows that $Q_j(\beta_j)$ is strictly convex for any $j = 1, \cdots, p'$, whenever $\tau + 1/\gamma < 1/2$.
\label{prop:conv}
\end{proposition}
\noindent In words, this shows that a sufficiently small choice of $\tau+1/\gamma$ is needed to ensure some form of convexity for the objective $Q(\boldsymbol{\beta})$. The first part of this proposition shows a unique global minimum exists for $Q(\boldsymbol{\beta})$ when $\tau + 1/\gamma < {\lambda_{min}(\bm{X}^T\bm{X})}/(2n)$. Such a result is quite restrictive, because it applies only to the low-dimensional setting of $n \leq p'$, where $\lambda_{min}(\bm{X}^T\bm{X})$ is strictly positive. The second part guarantees the coordinate-wise objective $Q_j(\beta_j)$ is strictly convex whenever $\tau + 1/\gamma  < 1/2$, a result which holds in the high-dimensional setting of $n > p'$. This coordinate-wise convexity is important for deriving the threshold function below.

For a main effect $J$, consider now its coordinate-wise minimization:
\begin{equation}
\min_{\beta_j} Q_j(\beta_j) = \min_{\beta_j} \left[ \frac{1}{2n}\| \bm{r}_{-j} - \bm{x}_j\beta_j\|^2_2 + \eta_{\lambda_s,\tau}\left\{ \|\boldsymbol{\beta}_{\mathcal{S}(j)}\|_{\lambda_s,\gamma} \right\} + \eta_{\lambda_c,\tau}\left\{ \|\boldsymbol{\beta}_{\mathcal{C}(j)}\|_{\lambda_c,\gamma} \right\}\right],
\end{equation}
where $\bm{r}_{-j} = \bm{y} - \bm{X}\boldsymbol{\beta} + \bm{x}_j\beta_j$ is the residual vector fitted without $\bm{x}_j$. Similarly, {for a} CME $J|K+$, its coordinate-wise minimization becomes:
\begin{equation}
\min_{\beta_{j|k+}} Q_{j|k+}(\beta_{j|k+}) = \min_{\beta_{j|k+}} \left[\frac{1}{2n}\| \bm{r}_{-(j|k+)} - \bm{x}_{j|k+}\beta_{j|k+}\|^2_2 + \eta_{\lambda_s,\tau}\left\{ \|\boldsymbol{\beta}_{\mathcal{S}(j)}\|_{\lambda_s,\gamma} \right\} + \eta_{\lambda_c,\tau}\left\{ \|\boldsymbol{\beta}_{\mathcal{C}(k)}\|_{\lambda_c,\gamma} \right\} \right].
\end{equation}
An optimization technique called \textit{majorization-minimization} (MM, see Chapter 12 of \citealp{Lan2010}) can now be used to derive a threshold function. The main idea of MM is as follows. Instead of minimizing the original objective function, one first obtains a {majorizing} surrogate function which lies above the desired objective. This surrogate is then minimized in place of the original objective. Under certain conditions, the solution iterates generated by repeating this procedure converge to a minimizer for the original problem \citep{Lan2010}. For $Q_j$ and $Q_{j|k+}$, a simple first-order expansion yields a nice majorizing surrogate function which can be minimized in closed form, as the following theorem demonstrates:

\begin{figure}[t]
\centering
\includegraphics[width=1.05\textwidth]{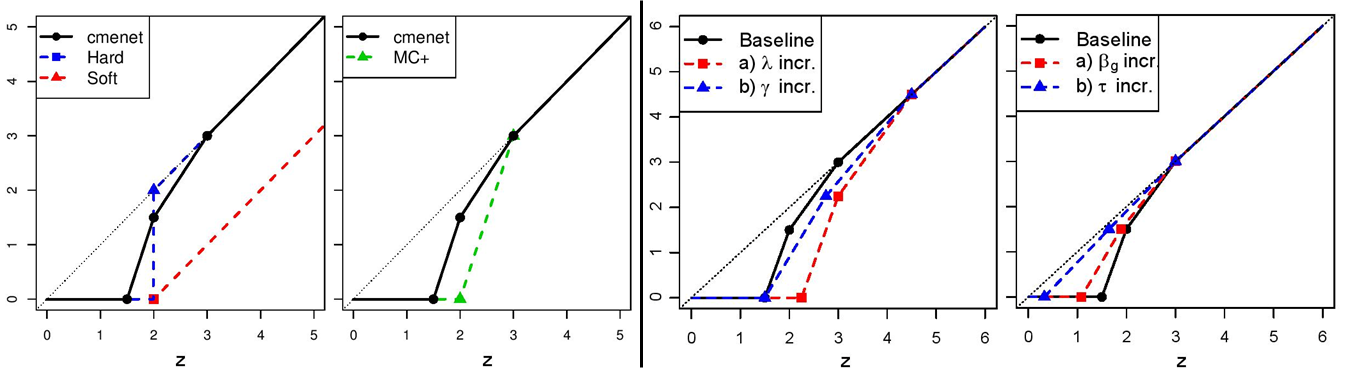}
\caption{(1st and 2nd plots) A comparison of the baseline threshold function $S_{\lambda_1,\lambda_2}$ (baseline setting: $(\lambda_1, \lambda_2, \gamma,\tau) = (1,0.5,3,0.05)$ with no selected group effects) with soft-, hard- and MC+ thresholding. (3rd plot) A comparison of the baseline threshold function with two new settings $(1.5,0.75,3.0.05)$ and $(1,0.5,4.5,0.05)$, all with no selected group effects. (Last) A comparison of the baseline threshold with two new settings $(1,0.5,3,0.05)$ and $(1,0.5,3,0.25)$, the latter with grouped norms $\|\beta_g\|_{\lambda_1,\gamma} = \|\beta_g\|_{\lambda_2,\gamma} = 5$.}
\label{fig:pen}
\end{figure}

\begin{theorem}[Threshold function]
Suppose $\tau + 1/\gamma < 1/2$. For fixed $\tilde{\boldsymbol{\beta}} \in \mathbb{R}^{p'}$, define $\bar{Q}_j(\cdot|\tilde{\boldsymbol{\beta}})$ and $\bar{Q}_{j|k+} (\cdot|\tilde{\boldsymbol{\beta}})$ as:
\begin{align*}
\begin{split}
\bar{Q}_j (\beta_j|\tilde{\boldsymbol{\beta}}) &= \frac{1}{2n}\| \bm{r}_{-j} - \bm{x}_j\beta_j\|^2_2 + \eta_{\lambda_s,\tau}\left\{ \| \tilde{\boldsymbol{\beta}}_{\mathcal{S}(j)}\|_{\lambda_s,\gamma} \right\} + \eta_{\lambda_c,\tau}\left\{ \|\tilde{\boldsymbol{\beta}}_{\mathcal{C}(j)}\|_{\lambda_c,\gamma} \right\}\\
& \quad + \tilde{\Delta}_{\mathcal{S}(j)} \left\{ g_{\lambda_s,\gamma}(\beta_j) - g_{\lambda_s,\gamma}(\tilde{\beta_j}) \right\} + \tilde{\Delta}_{\mathcal{C}(j)} \left\{  g_{\lambda_c,\gamma}(\beta_j) - g_{\lambda_c,\gamma}(\tilde{\beta_j}) \right\}, \text{ and}\\
\bar{Q}_{j|k+} (\beta_{j|k+}|\tilde{\boldsymbol{\beta}}) &= \frac{1}{2n}\| \bm{r}_{-({j|k+})} - \bm{x}_{j|k+}\beta_{j|k+}\|^2_2 + \eta_{\lambda_s,\tau}\left\{ \| \tilde{\boldsymbol{\beta}}_{\mathcal{S}(j)}\|_{\lambda_s,\gamma} \right\} + \eta_{\lambda_c,\tau}\left\{ \|\tilde{\boldsymbol{\beta}}_{\mathcal{C}(k)}\|_{\lambda_c,\gamma} \right\}\\
&  + \tilde{\Delta}_{\mathcal{S}(j)} \left\{  g_{\lambda_s,\gamma}(\beta_{j|k+}) - g_{\lambda_s,\gamma}(\tilde{\beta}_{j|k+}) \right\} + \tilde{\Delta}_{\mathcal{C}(k)} \left\{  g_{\lambda_c,\gamma}(\beta_{j|k+}) - g_{\lambda_c,\gamma}(\tilde{\beta}_{j|k+}) \right\},
\end{split}
\end{align*}
where $\tilde{\cdot}$ indicates the quantity is computed with $\tilde{\boldsymbol{\beta}}$ instead of $\boldsymbol{\beta}$. Then:
\ben[a)]
\item $\bar{Q}_j(\cdot|\tilde{\boldsymbol{\beta}})$ and $\bar{Q}_{j|k+}(\cdot|\tilde{\boldsymbol{\beta}})$ are majorization functions for ${Q}_j(\cdot)$ and $Q_{j|k+}(\cdot)$, respectively,
\item The unique minimizers of $\bar{Q}_j(\cdot|\tilde{\boldsymbol{\beta}})$ and $\bar{Q}_{j|k+}(\cdot|\tilde{\boldsymbol{\beta}})$ are given by $S_{\lambda_s,\lambda_c}(\bm{x}_j^T\bm{r}_{-j}/n;\tilde{\Delta}_{\mathcal{S}(j)},\tilde{\Delta}_{\mathcal{C}(j)})$ and $S_{\lambda_s,\lambda_c}(\bm{x}_{j|k+}^T\bm{r}_{-j|k+}/n;\tilde{\Delta}_{\mathcal{S}(j)},\tilde{\Delta}_{\mathcal{C}(k)})$, respectively. Here, $S_{\lambda_1,\lambda_2}(\cdot;\Delta_1, \Delta_2)$ is the \textup{threshold function}:
\een
\begin{equation}
\small
S_{\lambda_1,\lambda_2}(z;\Delta_1, \Delta_2) = 
\begin{dcases} 
z & \hspace{-0.4\textwidth} \textup{if } z \in [\lambda_{(1)}\gamma, \infty),\\
\textup{sgn}(z) \left(|z| - \Delta_{(1)} \right) / \left( {1 - \frac{\Delta_{(1)}}{\lambda_{(1)} \gamma}} \right)  & \\
& \hspace{-0.4\textwidth} \textup{if } z \in \Bigg[\lambda_{(2)}\gamma + \Delta_{(1)} \left( 1 - \frac{\lambda_{(2)}}{\lambda_{(1)}} \right), \lambda_{(1)}\gamma \Bigg),\\
\textup{sgn}(z) \left(|z| - \Delta_{(1)} - \Delta_{(2)} \right) / \left(1 - \frac{\Delta_{(1)}}{\lambda_{(1)} \gamma} - \frac{\Delta_{(2)}}{\lambda_{(2)} \gamma} \right) &\\
& \hspace{-0.4\textwidth} \textup{if } z \in \Bigg[\Delta_{(1)} + \Delta_{(2)}, \lambda_{(2)}\gamma + \Delta_{(1)} \left( 1 - \frac{\lambda_{(2)}}{\lambda_{(1)}} \right) \Bigg),\\
0, & \hspace{-0.4\textwidth} \textup{otherwise}.
\end{dcases}
\label{eq:thresh}
\end{equation}
\normalsize
where $\lambda_{(1)} = \max(\lambda_1,\lambda_2)$ and $\lambda_{(2)} = \min(\lambda_1,\lambda_2)$, with $\Delta_{(1)}$ and $\Delta_{(2)}$ its corresponding slopes.
\label{thm:thresh}
\end{theorem}

To better understand the shrinkage behavior of this new threshold function, the left two plots in Figure \ref{fig:pen} show a baseline setting of the \texttt{cmenet} threshold $S_{\lambda_1, \lambda_2}(z;\Delta_1,\Delta_2)$, compared with the soft-threshold function (corresponding to the shrinkage operator in LASSO), the hard-threshold function (corresponding to best-subset selection; see \citealp{Fea2001}), and the MC+ threshold function \citep{Mea2011}. The baseline setting for the proposed threshold $S_{\lambda_1, \lambda_2}(z;\Delta_1,\Delta_2)$ is set as $(\lambda_1, \lambda_2, \gamma,\tau) = (1, 0.5, 3,0.05)$, with $\|\boldsymbol{\beta}_g\|_{\lambda_1,\gamma} = \|\boldsymbol{\beta}_g\|_{\lambda_2,\gamma} = 0$ (i.e., no selected grouped effects). We see that the proposed threshold function is continuous and piecewise linear in four segments. Beginning from the left, the first segment is a horizontal line at zero, and represents the inner-product values for which a coefficient is shrunk to zero after thresholding. The last segment, which matches the identity line, represents the values for which the full coefficient signal is retained without any shrinkage. The middle two segments provide a two-step transition between these two extremes, with slopes controlled by the sibling and cousin penalties. Similar to the MC+ threshold, the \texttt{cmenet} threshold bridges the gap between the two extremes of full shrinkage and no shrinkage; however, the former threshold accomplishes this transition in one step, while the latter achieves this in two steps. This two-step transition for \texttt{cmenet} is a consequence of the two-tiered coupling effect from sibling and cousin groups.


Consider next the right two plots of Figure \ref{fig:pen}, which investigate the sensitivity of the proposed threshold $S_{\lambda_1, \lambda_2}(z;\Delta_1,\Delta_2)$ to changes in penalty parameters. From the first plot, an increase in $\lambda_1$, $\lambda_2$ or $\gamma$ appears to yield greater shrinkage of the coefficient signal. This is expected, because a larger choices of $\lambda_1$ and $\lambda_2$ induce greater regularization, and a larger $\gamma$ generates a ``more convex'' penalty (see \citealp{Mea2011}). From the second plot, an increase in the coupling parameter $\tau$ in the presence of selected group effects appears to greatly reduce signal shrinkage. This observation nicely demonstrates the earlier CME coupling principle in Section \ref{sec:couple}, where the selection of sibling or cousin effects increases the chances of a CME entering the model.

\subsubsection{Algorithm statement}
\label{sec:alg}

\begin{algorithm}[t]
\caption{\texttt{cmenet}: An algorithm for bi-level CME selection}
\label{alg:cmenet}
\begin{algorithmic}[1]
\Function{\texttt{cmenet}}{$\bm{X},\bm{y},\lambda_s, \lambda_c,\gamma,\tau, \boldsymbol{\beta} = \bm{0}_{p'}$} \Comment{Assume columns of $\bm{X}$ are normalized}
\stb Initialize $\bm{r} \leftarrow \bm{y} - \bar{\bm{y}}$, $\Delta_{\mathcal{S}(j)} = \lambda_s$, $\Delta_{\mathcal{C}(j)} = \lambda_c$ for $j = 1, \cdots, p$
\Repeat
\For{$j = 1, \cdots, p$} \Comment{For all main effects...}
\stb $\beta_0 \leftarrow \beta_j$, $\beta_j \leftarrow S_{\lambda_s,\lambda_c}\{ \bm{x}_j^T\bm{r}/n + \beta_0;\Delta_{\mathcal{S}(j)}, \Delta_{\mathcal{C}(j)}\}$ \Comment{Shrinkage}
\stb $\bm{r} \leftarrow \bm{r} + \bm{x}_j(\beta_0-\beta_j)$ \Comment{Update residual}
\stb $\Delta_{\mathcal{S}(j)} \leftarrow \Delta_{\mathcal{S}(j)} \exp\{-{\tau}/{\lambda_s}\left[ g_{\lambda_s,\gamma}(\beta_j) - g_{\lambda_s,\gamma}(\beta_0) \right]\}$ \Comment{Update slopes}
\stb $\Delta_{\mathcal{C}(j)} \leftarrow \Delta_{\mathcal{C}(j)} \exp\{-{\tau}/{\lambda_c}\left[ g_{\lambda_c,\gamma}(\beta_j) - g_{\lambda_c,\gamma}(\beta_0) \right]\}$
\EndFor
\For {$j = 1, \cdots, p$ and $k = 1, \cdots, p$} \Comment{For all CMEs (both $J|K+$ and $J|K-$) ...}
\stb $\beta_0 \leftarrow \beta_{j|k+}$, $\beta_{j|k+} \leftarrow S_{\lambda_s,\lambda_c}\{ \bm{x}_{j|k+}^T\bm{r}/n + \beta_0 ;\Delta_{\mathcal{S}(j)}, \Delta_{\mathcal{C}(k)} \}$ \Comment{Shrinkage}
\stb $\bm{r} \leftarrow \bm{r} + \bm{x}_{j|k+}(\beta_0-\beta_{j|k+})$ \Comment{Update residual}
\stb $\Delta_{\mathcal{S}(j)} \leftarrow \Delta_{\mathcal{S}(j)} \exp\{-{\tau}/{\lambda_s}\left[ g_{\lambda_s,\gamma}(\beta_{j|k+}) - g_{\lambda_s,\gamma}(\beta_0) \right]\}$ \Comment{Update slopes}
\stb $\Delta_{\mathcal{C}(k)} \leftarrow \Delta_{\mathcal{C}(k)} \exp\{-{\tau}/{\lambda_c}\left[ g_{\lambda_c,\gamma}(\beta_{j|k+}) - g_{\lambda_c,\gamma}(\beta_0) \right]\}$
\EndFor
\Until{$\boldsymbol{\beta}$ converges}\\
\Return the converged coefficient vector $\boldsymbol{\beta}$
\EndFunction
\end{algorithmic}
\end{algorithm}

Putting all the pieces together, Algorithm \ref{alg:cmenet} summarizes the detailed steps for \texttt{cmenet}, which minimizes the selection criterion $Q(\boldsymbol{\beta})$ given fixed parameters $\lambda_s$, $\lambda_c$, $\gamma$ and $\tau$. Starting with an initial solution of $\boldsymbol{\beta} = \bm{0}_{p'}$, the threshold function in \eqref{eq:thresh} is applied cyclically over each element in $\boldsymbol{\beta}$. This iterative procedure is then repeated until $\boldsymbol{\beta}$ converges. Using the majorization function in Theorem \ref{thm:thresh}, one can prove the convergence of \texttt{cmenet} to a stationary solution{.}
\begin{corollary}[Convergence of \texttt{cmenet}]
When $\tau + 1/\gamma < 1/2$, \textup{\texttt{cmenet}} converges to a stationary solution $\hat{\boldsymbol{\beta}}$ satisfying $\nabla Q(\hat{\boldsymbol{\beta}}) = 0$.
\label{cor:conv}
\end{corollary}
\noindent As for its running time, one can show that one coordinate descent cycle in \texttt{cmenet} over all $p'$ ME and CME coefficients requires $\mathcal{O}(np')$ work, because each coordinate descent step requires $\mathcal{O}(n)$ work. The linear running time in both sample size $n$ and total effects $p'$ is crucial for the computational efficiency of \texttt{cmenet}, particularly when a large number of main effects $p \gg 1$ is considered.

We mention here several extensions for \texttt{cmenet}. First, while Algorithm \ref{alg:cmenet} considers only the selection and estimation of CMEs, the proposed algorithm can easily be extended for the selection of both CMEs and \textit{other} covariate factors (whether continuous or discrete). For example, if the $l_1$-penalty were imposed on these latter factors, one can simply modify the coordinate descent loop in Algorithm \ref{alg:cmenet} by incorporating soft-threshold updates \citep{Don1995} to the coefficients of such factors. The algorithmic convergence for this extension is analogous to Corollary \ref{cor:conv}, and is not included for brevity. Second, we note that \texttt{cmenet}, as stated in Algorithm \ref{alg:cmenet}, is suitable for selecting binary CMEs -- CMEs which quantify the effect of a \textit{binary} factor at fixed levels of another factor, but not continuous CMEs -- CMEs which quantify the effect of a \textit{continuous} factor at fixed levels of another factor. One way to extend \texttt{cmenet} for the latter problem is to first (a) discretize the underlying continuous factor into two levels, then (b) perform \texttt{cmenet} on the resulting binary CMEs, and finally (c) quantify the continuous component of these continuous CMEs using the \textit{residuals} from \texttt{cmenet} as a new response vector. However, this extension requires further developments, and given the length of the current paper, we defer such an extension to future work. 

\subsection{Parameter tuning, warm starts and active set optimization}
\label{sec:comp}

While Algorithm \ref{alg:cmenet} provides an efficient method for minimizing the selection criterion $Q(\boldsymbol{\beta})$ given \textit{fixed} penalty parameters $\lambda_s$, $\lambda_c$, $\gamma$ and $\tau$, such parameters are typically not known in practice, and therefore require tuning. We present below a method for performing this tuning procedure, as well as two computational tools -- warm starts and active set optimization -- which greatly speed up this tuning in practice.


For parameter tuning, we adopt the relatively standard procedure (see, e.g., \citealp{Fea2001, Mea2011}) of finding the optimal penalty setting whose corresponding model (fitted using \texttt{cmenet}) minimizes some estimate of prediction error. In our implementation, called \texttt{cv.cmenet}\footnote{In later sections, the tuning procedure \texttt{cv.cmenet} is often referred to as simply \texttt{cmenet} for brevity.}, this prediction error is estimated using a technique called $K$-fold cross validation (or $K$-fold CV; see \citealp{Fea2001}), which randomly splits the observed data into $K$ parts, and uses one part of the data to validate the model fitted with the remaining $K-1$ parts. After obtaining this optimal penalty setting, the corresponding fitted model is then used for variable selection and prediction. For brevity, the specific details for \texttt{cv.cmenet} are summarized in Appendix \ref{sec:cvcmenet}.

One practical challenge for this tuning procedure is that there are four parameters ($\lambda_s, \lambda_c, \gamma, \tau)$ to tune for in \texttt{cv.cmenet}. Some guiding rules are therefore needed to efficiently explore this 4-d parameter space. The proposition below provides one such rule for ($\lambda_s, \lambda_c$):
\begin{proposition}[Search rule for $(\lambda_s,\lambda_c)$]
Suppose $\lambda_s + \lambda_c \geq \displaystyle \max_{j=1,\cdots,p'} | \bm{x}_j^T \bm{y} |/n$. When $Q(\boldsymbol{\beta})$ is strictly convex, the unique minimizer of $Q(\boldsymbol{\beta})$ is the zero solution $\boldsymbol{\beta} = \bm{0}_{p'}$.
\label{prop:maxl}
\end{proposition}
\noindent It should be noted that, in the high-dimensional setting of $n > p'$, $Q(\boldsymbol{\beta})$ cannot be strictly convex (see discussion for Proposition \ref{prop:conv}), so $\boldsymbol{\beta} = \bm{0}_{p'}$ is only a stationary solution. Nonetheless, the restriction of $\lambda_s + \lambda_c < \displaystyle \max_{j=1,\cdots,p'} | \bm{x}_j^T \bm{y}|/n$ allows for considerable reduction in the search for interesting choices of $\lambda_s$ and $\lambda_c$. From Proposition \ref{prop:conv}, another rule is $\tau + 1/\gamma < 1/2$, which ensures the strict convexity of the coodinate-wise problem and therefore the numerical stability of the optimization procedure. For brevity, the incorporation of these rules in \texttt{cv.cmenet} is outlined in Appendix \ref{sec:cvcmenet}.

Two computational tools can be used to greatly speed up the tuning procedure \texttt{cv.cmenet} in high-dimensions. The first tool, called \textit{warm starts}, makes use the converged solution from a previous parameter setting to initialize the optimization problem for the current setting. The use of warm starts in variable selection was popularized in \cite{Fea2007} for efficiently fitting multiple models along the full LASSO path, and we found such a tool to be equally effective for efficiently fitting multiple models over a grid of penalty parameters for \texttt{cmenet}. The second tool, called \textit{active set optimization} (see, e.g., \citealp{Mea2008, Fea2010}), performs coordinate descent updates over a small subset of \textit{active} variables, instead of over the full set of $p'$ variables. This technique is most effective when there are only a small number of active effects present, because one can avoid performing redundant coordinate descent updates on coefficients of inactive effects. Appendix \ref{sec:cvcmenet} provides specific details on how these two tools can be incorporated into \texttt{cv.cmenet}.

\subsection{CME screening rules}
\label{sec:screen}
When the number of main effects $p$ grows large, performing even one full coordinate descent over all $p' = p + 4{p \choose 2}$ total effects can be computationally cumbersome. One effective way of reducing computation time in such a situation is the use of screening rules, or \textit{strong rules}, which screen out a large number of inactive variables from consideration using previously-solved coefficient solutions. The term ``strong rules'' is first coined in \cite{Tea2012}, where the authors used previously-solved solutions along the LASSO path to screen out inactive effects for subsequent optimizations. We derive below similar strong rules for screening out inactive effects for \texttt{cmenet}, and reveal an interesting connection between these screening rules and CME coupling.

Suppose the parameters $\gamma$ and $\tau$ are fixed, and let $j$ index a variable of interest (ME or CME), with $\mathcal{S}$ and $\mathcal{C}$ its corresponding sibling and cousin group. Furthermore, let $\hat{\boldsymbol{\beta}}(\lambda_s,\lambda_c)$ be an optimal solution of the selection criterion $Q(\boldsymbol{\beta})$ under penalties $\lambda_s$ and $\lambda_c$, and let $c_j(\lambda_s,\lambda_c) = \bm{x}_j^T(\bm{y}-\bm{X}\hat{\boldsymbol{\beta}}(\lambda_s,\lambda_c))/n$ denote the inner-product of effect $j$ with the current residual vector. Denoting $\lambda_s^1 > \lambda_s^2 > \cdots > \lambda_s^L$ and $\lambda_c^1 > \lambda_c^2 > \cdots > \lambda_c^M$ as the desired (decreasing) penalty sequences for $\lambda_s$ and $\lambda_c$, the screening procedure can be summarized by the following three strong rules:
\ben
\item Suppose there are no active effects in $\mathcal{S}$ and $\mathcal{C}$ for penalty settings $(\lambda_s^{l-1},\lambda_c^{m})$ or $(\lambda_s^{l},\lambda_c^{m-1})$. Then effect $j$ is marked as \textit{inactive} for penalty setting $(\lambda_s^{l},\lambda_c^{m})$ if:
\begin{equation}
|c_j(\lambda_s^{l-1},\lambda_c^{m})| < \lambda_s^{l} + \lambda_c^m + \frac{\gamma}{\gamma-2} (\lambda_s^l - \lambda_s^{l-1}) \quad \text{or} \quad |c_j(\lambda_s^{l},\lambda_c^{m-1})| < \lambda_s^l + \lambda_c^{m} + \frac{\gamma}{\gamma-2} (\lambda_c^m - \lambda_c^{m-1}).
\label{eq:nosc}
\end{equation}
\item If there are no active effects in the \textit{sibling} group $\mathcal{S}$ for penalty setting $(\lambda_s^{l-1},\lambda_c^m)$, then effect $j$ is marked as \textit{inactive} for penalty setting $(\lambda_s^l,\lambda_c^m)$ if:
\begin{equation}
|c_j(\lambda_s^{l-1},\lambda_c^{m})| < \lambda_s^{l} + \Delta_{\mathcal{C}}' + \frac{\gamma}{\gamma-(\Delta_{\mathcal{C}}'/\lambda_c^m +1)} (\lambda_s^l - \lambda_s^{l-1}),
\label{eq:nos}
\end{equation}
where $\Delta_{\mathcal{C}}' = \lambda_c^{m} \exp\left\{ -\tau \|\boldsymbol{\beta}_{\mathcal{C}}(\lambda_s^{l-1},\lambda_c^m)\|_{\lambda_c^m,\gamma}/{\lambda_c^m} \right\}$.
\item If there are no active effects in the \textit{cousin} group $\mathcal{C}$ for penalty setting $(\lambda_s^{l}, \lambda_c^{m-1})$, then effect $j$ is marked as \textit{inactive} for penalty setting $(\lambda_s^l,\lambda_c^m)$ if:
\begin{equation}
|c_j(\lambda_s^{l},\lambda_c^{m-1})| < \Delta_{\mathcal{S}}' + \lambda_c^{m} + \frac{\gamma}{\gamma-(\Delta_{\mathcal{S}}'/\lambda_s^l +1)} (\lambda_c^m - \lambda_c^{m-1}),
\label{eq:noc}
\end{equation}
where $\Delta_{\mathcal{S}}' = \lambda_s^{l} \exp\left\{ -\tau \|\boldsymbol{\beta}_{\mathcal{S}}(\lambda_s^l,\lambda_c^{m-1})\|_{\lambda_s^l,\gamma}/{\lambda_s^l} \right\}$.
\een
\noindent A theoretical derivation of these rules is provided in Appendix \ref{sec:cmescreen}.

While these three rules may appear complicated and technical, they are in fact quite intuitive to understand. All three rules consider conditions under which it would be ``safe'' to screen out effect $j$ from the optimization problem for the penalty setting $(\lambda_s^l, \lambda_c^m)$. The first rule applies when there are no active effects in $\mathcal{S}$ and $\mathcal{C}$ from previous penalty settings, and screens out effect $j$ if the previous inner-products $c_j(\lambda_s^{l-1},\lambda_c^m)$ or $c_j(\lambda_s^l,\lambda_c^{m-1})$ are within the upper bounds provided in \eqref{eq:nosc}. The intuition here is that if effect $j$ is not correlated enough with the residual vectors at the previous penalty settings $(\lambda_s^{l-1},\lambda_c^m)$ or $(\lambda_s^l,\lambda_c^{m-1})$, then it cannot ``catch up'' in time to be active for the current setting $(\lambda_s^l,\lambda_c^m)$ (see \citealp{Tea2012} for details). This first rule can be viewed as an extension of the MC+ strong rule in \cite{LB2015} to the current model. The second rule applies when there are no active effects in the sibling group $\mathcal{S}$ (but some in cousin group $\mathcal{C}$) for the previous setting $(\lambda_s^{l-1},\lambda_c^m)$. In such a scenario, effect $j$ is screened out if the previous inner-product $c_j(\lambda_s^{l-1},\lambda_c^m)$ is within the upper bound in \eqref{eq:nos}. The key difference between this and the first rule is that, as more effects are selected in the cousin group $\mathcal{C}$, the linearized slope $\Delta_{\mathcal{C}}'$ decays smaller than $\lambda_c^m$, which then decreases the screening bound in \eqref{eq:nos} compared to the original bound in \eqref{eq:nosc}\footnote{Here, we assume the last term in both \eqref{eq:nosc} and \eqref{eq:nos} are nearly equal in this comparison; the discrepancy between \eqref{eq:nosc} and \eqref{eq:nos} is dominated by the first two terms for most feasible parameter settings.}. In other words, the presence of coupled cousin effects from a previous setting can \textit{decrease} the screening power of strong rules for the current setting. This is quite similar to the CME coupling phenomenon in Section 3.2, except instead of encouraging the \textit{selection} of effect $j$, the coupled CMEs make it more diffcult to \textit{screen out} effect $j$ via strong rules. The third rule, which applies when there are no previously-active cousins in $\mathcal{C}$ (but some siblings in $\mathcal{S}$), enjoys a similar interpretation: as more siblings are coupled in from $\mathcal{S}$ at a previous setting, effect $j$ becomes more difficult to screen out via strong rules.


Lastly, we note that while these three rules do screen out a large proportion of inert CMEs, it is possible (but highly unlikely) that an active CME is erroneously screened out. This is illustrated numerically in the following section. To prevent any false-negative screenings, we recommend that the KKT conditions (see equation \eqref{eq:majsolkkt} in the Appendix) be checked as a final step for each optimization problem.

\section{Simulations}
\label{sec:sim}

We now explore the performance of the proposed method in several simulation studies. Table \ref{tbl:sim} summarizes the test settings for these simulations, with varying sample sizes $n$ and main effects $p$, varying number of active groups $x$ and active effects within a group $y$ (denoted as G$x$A$y$), and whether the grouped effects are siblings or cousins (main effect models are considered here as well). Active effects are assigned a value of 1 in the coefficient vector $\boldsymbol{\beta}$, and non-active effects assigned a value of 0. Each simulation case is then replicated 100 times, with the model matrix $\bm{X}$ simulated from the equicorrelated latent model in Section \ref{sec:group} with $\rho = 0$ and $\rho = 1/\sqrt{2}$, and the response $\bm{y}$ simulated independently from $\mathcal{N}(\bm{X}\boldsymbol{\beta},\bm{I}_n)$. For brevity, we only report the results for $(n,p) = (50,50), (100,100)$ and $(150,150)$ with G4A2 and G6A2, but similar conclusions hold for other settings.

\begin{table}[t]
\centering
\begin{threeparttable}
\small
\begin{tabular}{ c || c }
\toprule
\textit{Simulation parameters} & \textit{Settings}\\
\toprule
Sample size  & $n = 50, \; 100 \text{ or } 150 $\\
\hline
\# of main effects considered & $p = 50, \; 100 \text{ or } 150 $\\
(total effects considered) & $\left( p' =  p + 4 {p \choose 2} = 4,950, \; 19,900 \text{ or } 44,850 \right)$ \\
\hline
\# of active groups & 6 \text{or} 8\\
\hline
\# of active effects within a group & $2 \text{ or } 3$ \\
\hline
Effect type & Siblings, cousins, main effects\tnote{1} \\
\hline
Latent correlation & $\rho = 0 \text{ or } 1/\sqrt{2}$\\
\toprule
\end{tabular}
\caption{Test settings for simulation study.}
\label{tbl:sim}
\normalsize
\begin{tablenotes}
\item[1] \footnotesize{\# of active MEs is set as \# of active groups.}
\end{tablenotes}
\end{threeparttable}
\end{table}

Under such a set-up, our simulations aim to answer two questions: (a) Does the proposed method \texttt{cmenet} yield improved selection of CMEs compared to more generic selection methods? (b) For an active CME, say $J|K+$, is \texttt{cmenet} more effective at identifying this conditional, non-additive relation between $J$ and $K$, compared to the more traditional 2FI analysis? To answer the first question, we compare \texttt{cmenet} with two generic variable selection techniques from the literature: the LASSO \citep{Tib1996} using the R package \textsc{glmnet} \citep{Fea2009}, and \texttt{SparseNet} \citep{Mea2011} using the R package \textsc{sparsenet} \citep{Mea2012}. All three methods perform selection on the \textit{same} set of MEs and CMEs, with penalty parameters tuned using $10$-fold CV. In this comparison, a better selection performance for \texttt{cmenet} shows that the proposed penalty $Q(\boldsymbol{\beta})$ is more appropriate for selecting CMEs compared to generic penalties. To answer the second question, we compare \texttt{cmenet} with a popular selection method called \texttt{hierNet} \citep{Bea2013} for selecting 2FIs. A better selection performance for \texttt{cmenet} over \texttt{hierNet} thereby demonstrates the effectiveness of the proposed method in identifying the conditional, non-additive nature of CMEs.

We employ two criteria to conduct the above comparisons. The first criterion returns the number of misspecified variables: $\#\{\mathcal{A} \setminus \hat{\mathcal{A}}_n\} + \#\{\hat{\mathcal{A}}_n \setminus \mathcal{A}\}$, where $\mathcal{A}$ is the true active set of MEs and CMEs, and $\hat{\mathcal{A}}_n$ is the set of selected effects after $n$ observations. Smaller values of this indicate better selection performance. Such a criterion is appropriate for \texttt{cmenet}, LASSO and \texttt{SparseNet}, which perform selection on the MEs and CMEs in $\mathcal{A}$, but a slight modification is needed for \texttt{hierNet}, which performs selection on the traditional 2FIs. To this end, let $\mathcal{A}^{(ME)}$ consist of the original MEs in active set $\mathcal{A}$ as well as the parent MEs of the CMEs in $\mathcal{A}$, and let $\mathcal{A}^{(2FI)}$ consist of the 2FIs corresponding to the CMEs in $\mathcal{A}$. The misspecification criterion for \texttt{hierNet} can then be written as: $\#\{\mathcal{A}^{(ME)} \setminus \hat{\mathcal{A}}_n^{(ME)}\} + \#\{ \hat{\mathcal{A}}_n^{(ME)} \setminus \mathcal{A}^{(ME)} \} + \#\{ \mathcal{A}^{(2FI)} \setminus \hat{\mathcal{A}}_n^{(2FI)}\} + \#\{\hat{\mathcal{A}}_n^{(2FI)} \setminus \mathcal{A}^{(2FI)}\}$, where $\hat{\mathcal{A}}_n^{(ME)}$ and $\hat{\mathcal{A}}_n^{(2FI)}$ are the selected MEs and 2FIs from \texttt{hierNet}. Put another way, this modified criterion first translates the true CME model into its component MEs and 2FIs (see the identities in \eqref{eq:cmeconstr}), then reports the number of misspecifications for the fitted \texttt{hierNet} model based on these component effects. The second criterion is the mean-squared prediction error (MSPE): $\mathbb{E}\|\bm{y}_{new} - \bm{X}_{new}\hat{\boldsymbol{\beta}}\|_2^2$, where $(\bm{X}_{new},\bm{y}_{new})$ is an out-of-sample dataset with $n_{new} = 20$ observations simulated from the true model $\mathcal{A}$. Smaller MSPE values suggest better predictive performance. Here, the focus is on a method which yields the best \textit{selection} performance of CMEs (first criterion); however, such a method should have comparable \textit{predictive} performance to other methods (second criterion).

\begin{sidewaystable}[p]
\centering
\includegraphics[width=1.05\textwidth]{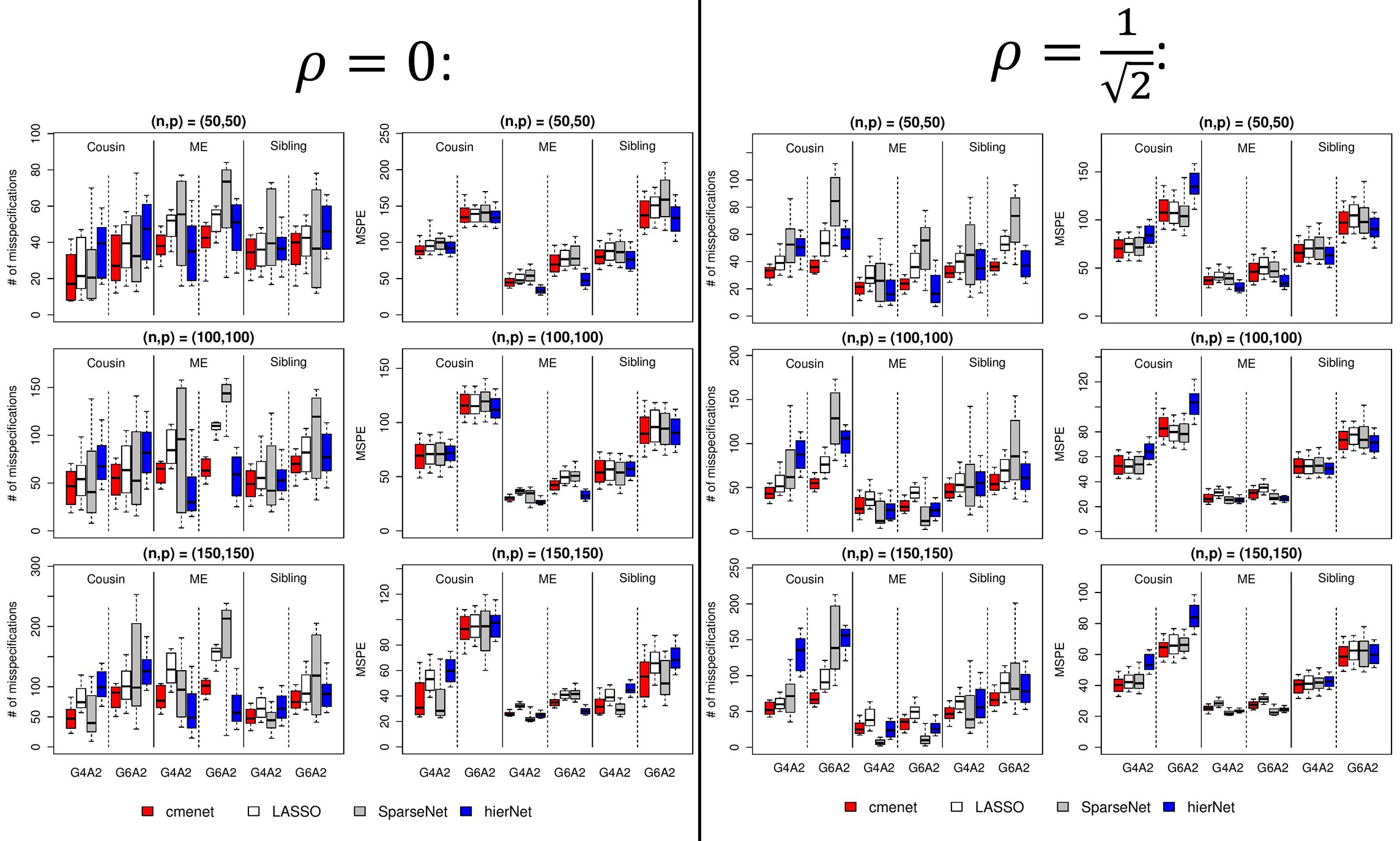}
\captionof{figure}{Boxplots of the 10\%, 25\%, 50\%, 75\% and 90\% quantiles for the \# of misspecifications and MSPE, with $(n,p)=(50,50)$ (top), $(n,p)=(100,100)$ (middle) and $(n,p)=(150,150)$ (bottom), using a latent correlation of $\rho=0$ (left) and $\rho = 1/\sqrt{2}$ (right).}
\label{fig:sim}
\end{sidewaystable}

Figures \ref{fig:sim} show the number of misspecifications and MSPE for the four methods with $\rho = 0$ and $\rho = 1/\sqrt{2}$, under the simulation settings presented earlier. Consider first the sibling and cousin models in the $\rho=0$ setting (left part of Figure \ref{fig:sim}), where the underlying MEs are uncorrelated. For these models, \texttt{cmenet} provides noticeably improved selection performance over LASSO and \texttt{SparseNet} for nearly all simulation settings. This shows that the penalization scheme in $Q(\boldsymbol{\beta})$ is indeed more effective than generic penalties for selecting active CMEs; by accounting for the implicit group structure of CMEs, the proposed method can better guide the variable selection procedure using the novel principle of CME coupling. \texttt{cmenet} also yields a sizable selection improvement over \texttt{hierNet} for sibling and cousin models, which shows that the proposed approach can better identify the conditional, non-linear nature of CMEs compared to traditional 2FI analysis. One likely explanation is that, because a CME can be decomposed into its component ME and 2FI effects (recall the identities in \eqref{eq:cmeconstr} and Rule 1 of \citealp{SW2016}), the selection signal of an active CME is much stronger than the signals from its component ME and 2FI effects. \texttt{cmenet}, by performing selection directly on the CMEs with greater signal, can more easily identify the underlying active effects compared to \texttt{hierNet}, which performs selection on its component ME and 2FI effects with diluted signals. As for MSPE, \texttt{cmenet} enjoys comparable or improved performance to the other three methods, which is as desired.

Consider next the main effect models for $\rho = 0$ (left part of Figure \ref{fig:sim}). We see that \texttt{cmenet} enjoys superior selection performance to LASSO and \texttt{SparseNet}, which demonstrates the effectiveness of the CME reduction principle in reducing selected CMEs into its underlying parent ME. Compared to \texttt{hierNet}, \texttt{cmenet} provides comparable (but slightly worse) selection for these main effect models, an observation not too surprising given that the proposed method specifically tackles the problem of CME selection. \texttt{cmenet} is therefore most effective in applications where one expects some conditional effects to be active in the model; in other words, in applications where CMEs represent interpretable, domain-specific phenomena. 

Finally, consider the results for $\rho = 1/\sqrt{2}$ (right part of Figure \ref{fig:sim}), where the underlying MEs are moderately correlated. For the sibling and cousin models, \texttt{cmenet} again provides an improvement in selection performance over the other three methods, with this improvement much greater than that for the uncorrelated setting $\rho=0$. Such an observation is expected in light of Section \ref{sec:group}, because the CME group structure is most prominent for moderate choices of $\rho$. For the main effect models, \texttt{cmenet} and \texttt{hierNet} again provide the best selection performance, with the relative performance of \texttt{cmenet} noticeably better than that for $\rho=0$. This again can be explained by the more pronounced CME group structure for moderate $\rho$, which allows for more effective CME reduction. As before, the MSPE for \texttt{cmenet} is comparable to or better than the other three methods, which is as desired.

\begin{figure}[t]
\centering
\includegraphics[width=0.8\textwidth]{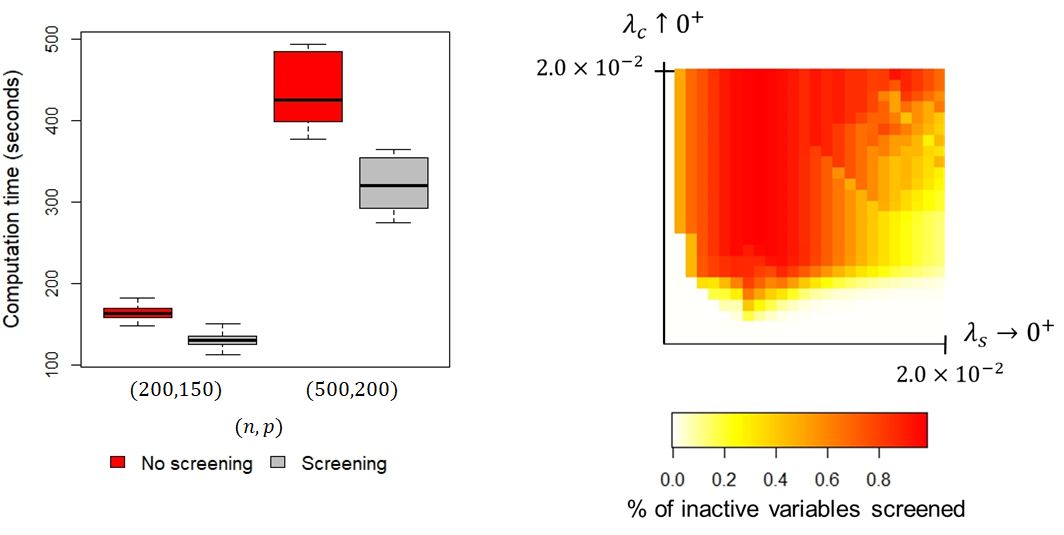}
\caption{(Left) Boxplots of computation times for \textup{\texttt{cmenet}} with $(n,p) = (200,150)$ and $(500,200)$; (Right) Proportion of inactive variables screened for $(n,p) = (200,150)$.}
\label{fig:screen}
\end{figure}

To numerically demonstrate the effectiveness of the CME screening rules in Section \ref{sec:screen}, the left plot in Figure \ref{fig:screen} shows the boxplots of the computation times for \texttt{cmenet} with $(n,p) = (200,150)$ and $(500,200)$, under a G2A6 sibling model with latent correlation $\rho = 0$. We see that the proposed screening rules significantly reduce computation time, with over 20\% reduction in median time for $(n,p) = (200,150)$, and 30\% reduction for $(n,p) = (500,200)$. This effectiveness appears to grow for larger sample sizes $n$ and greater number of main effects $p$, which is as desired. The right plot in Figure \ref{fig:screen} shows the proportion of inactive variables removed by the screening procedure for $(n,p) = (200,150)$ as a function of the sibling and cousin penalties $\lambda_s$ and $\lambda_c$. We see that the proposed screening rules correctly remove a large proportion of inactive variables (over 80\% for smaller $\lambda_s$ and $\lambda_c$), which greatly speeds up the ensuing coordinate descent algorithm. In total, only 3 active variables were incorrectly screened over all values of $(\lambda_s,\lambda_c)$ tested, and all such violations were corrected in post-convergence check of KKT conditions.

\section{Polygenic association study on fly wing shape}
\label{sec:appl}

In this section, we demonstrate the usefulness of \texttt{cmenet} for an important, real-world problem on polygenic association. Polygenes are a group of non-epistatic genes which serve as biological markers for many characteristics of interest called phenotypes (e.g., susceptibility to diabetes for youth \citep{Rea1999} and major depressive disorders \citep{Dea2015}), and the association of influential polygenes to particular phenotypes is an important area of research in the biomedical community. Here, we investigate the polygenic association for the wing shape of \textit{Drosophila Melanogaster}, the common fruit fly.



The data employed here is collected from a study by \cite{Wea2001}, where the authors considered $p = 48$ homozygous (i.e., binary\footnote{For organisms with diploid cells (including \textit{Drosophila Melanogaster}), chromosomes are found in pairs; these chromosome pairs can be further categorized as either \textit{heterozygous} -- meaning the pair contains different alleles for each gene, or \textit{homozygous} -- meaning the pair contains identical alleles for each gene. For dominant (+) and recessive (--) alleles, heterozygous pairs allow for four allele combinations (+,+), (+,--), (--,+) and (--,--), while homozygous pairs allow for two combinations (+,+) and (--,--). For this fly wing study, \cite{Wea2001} found very little heterozygous behavior on chromosome 2, and reported subsequent results using modified homozygous chromosomes, which are binary and fit within the framework of this paper.}) polygene markers on the second chromosome of \textit{Drosophila Melanogaster} and its effect on fly wing shape, using $n = 701$ observations collected from recombinant isogenic lines. The response of interest is a continuous index for wing shape, which incorporates both the width of the wing across the middle and the width across the base. As in simulation studies, our focus lies primarily on the \textit{selection} of important CMEs, which here represents the effect of a gene conditional on another gene being active or absent. This is because the identification of these novel conditional effects yields valuable insight into the activation structure of gene-gene interactions, whereas the more traditional two-factor interaction analysis can be less interpretable in such a setting.

Here, we compare the analysis provided by \texttt{cmenet} with that from \texttt{hierNet}. As before, \texttt{cmenet} performs selection on MEs and CMEs ($p' = p + 4 {p \choose 2} = 4,560$ variables in total), while \texttt{hierNet} performs MEs and 2FIs ($p'' = p + {p \choose 2} = 1,176$ variables in total). The purpose of such a comparison is to understand the practical advantages and disadvantages in employing the novel CMEs as basis functions, compared to the typical approach of using 2FIs for analyzing gene-gene interactions \citep{Cor2009}. For brevity, we do not include either the LASSO or \texttt{SparseNet} selection of CMEs in this comparison, because it was already shown in Section \ref{sec:sim} that \texttt{cmenet} enjoys better selection performance.


\begin{table}[t]
\centering
\begin{tabular}{ c || c | c c }
\toprule
\textit{Method} & \textit{\# of selected effects}  & \multicolumn{2}{c}{\textit{Some selected effects (p-values)}}\\
\toprule
\texttt{cmenet} & 21 & $\text{g}14|\text{g}27$- $(6.1 \times 10^{-4})$, & \multirow{4}{*}{$\text{g}45|\text{g}10$+ $(7.3 \times 10^{-7})$}\\
 & & $\text{g}14|\text{g}38$+ $(2.0 \times 10^{-2})$, & \\
 & & $\text{g}17|\text{g}14$- $(1.6 \times 10^{-12})$, & \\
 & & $\text{g}23|\text{g}14$+ $(2.5 \times 10^{-30})$ & \\
\hline
\texttt{hierNet} & 129 & \multirow{2}{*}{$\text{g}14 \; (8.3 \times 10^{-1})$} & $\text{g}45 \; (1.5 \times 10^{-1})$,\\
 & & & $\text{g}45*\text{g}10 \; (8.1 \times 10^{-1})$\\
\toprule
\end{tabular}
\captionof{table}{Number of selected effects and some selected effects (p-values bracketed) from \texttt{\textup{cmenet}} and \texttt{\textup{hierNet}} in the gene association study of fly wing shape.}
\label{tbl:wingsel}
\end{table}

\begin{figure}[t]
\centering
\includegraphics[width=0.4\textwidth]{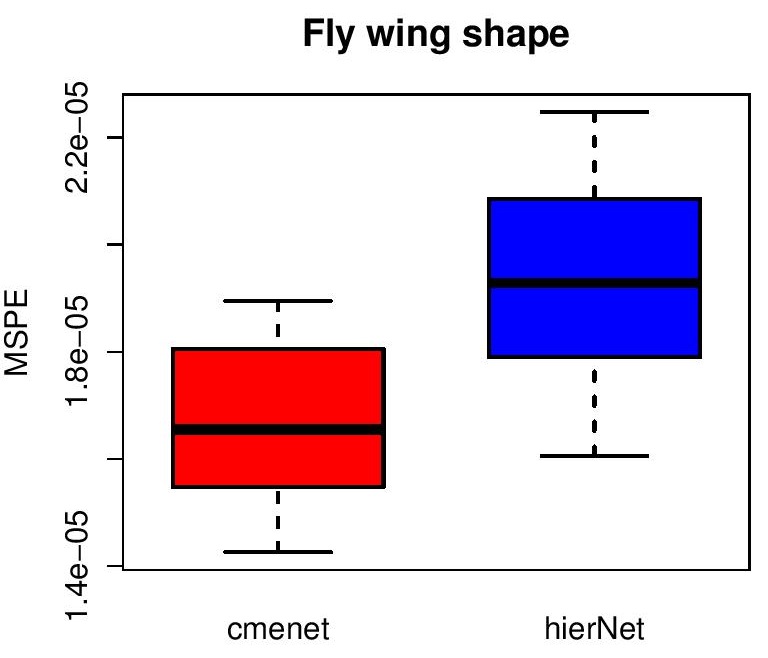}
\vspace{-0.5cm}
\captionof{figure}{Boxplots of the 10\%, 25\%, 50\%, 75\% and 90\% MSPE quantiles for \textup{\texttt{cmenet}} and \textup{\texttt{hierNet}} in the gene association study of fly wing shape.}
\label{fig:wingerr}
\end{figure}

Consider first Table \ref{tbl:wingsel}, which shows (a) the number of selected effects for \texttt{cmenet} and \texttt{hierNet}, and (b) some selected effects for each method, along with their corresponding p-values from a regular linear model fit. We see that the fitted model from \texttt{cmenet}, which has 21 selected effects, is much smaller than the model returned by \texttt{hierNet}, which has 129 selected effects. This model parsimony for \texttt{cmenet} suggests that there are indeed active CMEs for the problem at hand, i.e., there are certain polygenes which affect wing shape \textit{only} in the presence or absence of other polygenes. Taking a closer look at some of the selected effects for \texttt{cmenet} and \texttt{hierNet} from Table \ref{tbl:wingsel}, two interesting insights can be observed on this conditional gene association structure. From the first column of selected effects, \texttt{hierNet} deemed the 14-th polygene g$14$ to be active, while \texttt{cmenet} instead selected the two sibling effects g$14|$g$27$- and g$14|$g$38$+, and the two cousin effects g$17|$g$14$- and g$23|$g$14$+. In other words, under traditional analysis, gene g14 is deemed influential in all situations, whereas the conclusion is more nuanced under the proposed CME analysis, with g14 influential (a) when gene g27 is absent or gene g38 is active, or (b) in inhibiting gene g17 or activating gene g23. The latter provides a more careful analysis of the signal from g14, and judging by the much smaller p-values for these conditional effects, also yields greater insight on the underlying gene activation structure. From the second column of selected effects, \texttt{hierNet} deemed both g45 and its interaction g45$*$g10 to be active, while \texttt{cmenet} selected only the CME g45$|$g10+. This nicely illustrates why \texttt{cmenet} provides parsimonious models: by selecting the CME g45$|$g10+ in place of its component ME g45 and 2FI g45$*$g10, we obtain a smaller model with considerably smaller p-values, which is as desired (this is akin to Rule 1 of \citealp{SW2016} for selecting CMEs in designed experiments; see Section \ref{sec:cmegroups}, especially equation \eqref{eq:cmeconstr}).

Consider next Figure \ref{fig:wingerr}, which shows the MSPE boxplots for \texttt{cmenet} and \texttt{hierNet} in predicting the continuous wing shape index. Here, MSPE is estimated by randomly sampling 80\% of the data for model training, then using the remaining 20\% to test the trained model; this procedure is then repeated 200 times to provide error variability. We see that \texttt{cmenet} enjoys considerable improvements over \texttt{hierNet} in terms of MSPE, yielding at least a 12\% reduction at all five error quantiles. This again reaffirms the likely conditional nature of the underlying polygenic association structure, with certain polygenes active only in the presence or absence of other polygenes.

To summarize, this gene association study highlights two important advantages of \texttt{cmenet}. First, in applications where CMEs are interpretable phenomena, the proposed selection method can provide much more parsimonious models compared to traditional analysis using two-way interactions, and can yield greater insight on the underlying problem of interest. This is particularly true in genetic applications, where selected CMEs can be used to further investigate \textit{why} some genes are conditionally active, and \textit{why} some play a more supportive role in \textit{activating} or \textit{inhibiting} other genes. Second, when CMEs have natural domain-specific interpretations, using such effects as basis functions can also improve the predictive performance of the fitted model as well.

\section{Conclusion and future work}

In this paper, a new method is presented for selecting binary variables and a set of reparametrized variables called conditional main effects (CMEs) from observation data. While CMEs are intuitive basis functions with appealing interpretations in many applications, existing selection methods can perform poorly due to the inherent grouped structure of these effects. We proposed a novel selection method called \texttt{cmenet}, which accounts for this underlying grouped structure using two selection principles called CME coupling and reduction; the former allows CMEs to more easily enter the model given selected siblings or cousins, and the latter encourages the selection of the underlying ME given many selected siblings or cousins. A coordinate descent algorithm is then introduced for minimizing the selection criterion, and several computational tools are proposed for efficient optimization and parameter tuning in high-dimensions. Simulation studies showed considerable improvements for \texttt{cmenet} over existing methods with respect to selection accuracy. Applied to a real-world gene association study on fly wing shape, the proposed method provides not only improved predictive performance over the standard two-way interaction analysis, but also a more parsimonious and interpretable model which reveals important insights on gene activation behavior.

Given the positive results here, there are many exciting avenues for future work. First, in the high-dimensional setting of $p \gg 1$, the tuning of the four selection parameters in $Q(\boldsymbol{\beta})$ can be computationally expensive due to the grid structure of feasible parameter combinations in \texttt{cv.cmenet}. With recent advances on the topic of optimal designs for convex spaces (e.g., \citealp{LJ2015, MJ2016}), it may be interesting to see whether the use of such designs as candidate settings allows for more efficient parameter tuning. Second, we are working to broaden the proposed methodology to higher-order conditional effects, e.g., the effect of $A$ conditional on both $B+$ and $C+$. The main challenge here is again computational efficiency, but such a direction would enable the investigation of, say, more complex activation phenomena in the earlier gene study. Lastly, we are interested in extending the current framework for selecting the continuous CMEs mentioned earlier in Section \ref{sec:alg}. This would allow the proposed methodology to be applicable for more general datasets, and we look forward to exploring this in future research.

\noindent \textbf{Acknowledgements}

\noindent The authors gratefully acknowledge helpful advice from two referees, particularly one referee who pointed out a minor mistake in Proposition \ref{prop:conv}, and whose comments led to the development of the CME screening rules in Section \ref{sec:screen}. \if1\blind{The authors also thank Prof. Ben Haaland for his useful comments and suggestions.} \fi An efficient C++ implementation of \texttt{cmenet} and \texttt{cv.cmenet} is available in the R package \textsc{cmenet} in CRAN.

\bibliography{references}

\begin{thebibliography}{}

\bibitem[Bien et~al., 2013]{Bea2013}
Bien, J., Taylor, J., and Tibshirani, R. (2013).
\newblock A lasso for hierarchical interactions.
\newblock {\em The Annals of Statistics}, 41(3):1111--1141.

\bibitem[Breheny, 2015]{Bre2015}
Breheny, P. (2015).
\newblock The group exponential lasso for bi-level variable selection.
\newblock {\em Biometrics}, 71(3):731--740.

\bibitem[Breheny and Huang, 2009]{BH2009}
Breheny, P. and Huang, J. (2009).
\newblock Penalized methods for bi-level variable selection.
\newblock {\em Statistics and Its Interface}, 2(3):369.

\bibitem[Chari and Dworkin, 2013]{CD2013}
Chari, S. and Dworkin, I. (2013).
\newblock The conditional nature of genetic interactions: the consequences of
  wild-type backgrounds on mutational interactions in a genome-wide modifier
  screen.
\newblock {\em PLoS Genetics}, 9(8):e1003661.

\bibitem[Cordell, 2009]{Cor2009}
Cordell, H.~J. (2009).
\newblock Detecting gene--gene interactions that underlie human diseases.
\newblock {\em Nature Reviews Genetics}, 10(6):392--404.

\bibitem[De~Moor et~al., 2015]{Dea2015}
De~Moor, M.~H., Van Den~Berg, S.~M., Verweij, K.~J., Krueger, R.~F., Luciano,
  M., Vasquez, A.~A., Matteson, L.~K., Derringer, J., Esko, T., and Amin, N.
  (2015).
\newblock Meta-analysis of genome-wide association studies for neuroticism, and
  the polygenic association with major depressive disorder.
\newblock {\em JAMA Psychiatry}, 72(7):642--650.

\bibitem[Donoho, 1995]{Don1995}
Donoho, D.~L. (1995).
\newblock De-noising by soft-thresholding.
\newblock {\em IEEE Transactions on Information Theory}, 41(3):613--627.

\bibitem[Fan and Li, 2001]{FL2001}
Fan, J. and Li, R. (2001).
\newblock Variable selection via nonconcave penalized likelihood and its oracle
  properties.
\newblock {\em Journal of the American Statistical Association},
  96(456):1348--1360.

\bibitem[Finney, 1945]{Fin1945}
Finney, D. (1945).
\newblock The fractional replication of factorial arrangements.
\newblock {\em Annals of Eugenics}, 12:291--303.

\bibitem[Frank and Friedman, 1993]{FF1993}
Frank, I.~E. and Friedman, J.~H. (1993).
\newblock A statistical view of some chemometrics regression tools.
\newblock {\em Technometrics}, 35(2):109--135.

\bibitem[Friedman et~al., 2007]{Fea2007}
Friedman, J., Hastie, T., H{\"o}fling, H., and Tibshirani, R. (2007).
\newblock Pathwise coordinate optimization.
\newblock {\em The Annals of Applied Statistics}, 1(2):302--332.

\bibitem[Friedman et~al., 2001]{Fea2001}
Friedman, J., Hastie, T., and Tibshirani, R. (2001).
\newblock {\em The {E}lements of {S}tatistical {L}earning}.
\newblock Springer.

\bibitem[Friedman et~al., 2009]{Fea2009}
Friedman, J., Hastie, T., and Tibshirani, R. (2009).
\newblock \textsc{glmnet}: Lasso and elastic-net regularized generalized linear
  models.
\newblock {\em R package version 1}.

\bibitem[Friedman et~al., 2010]{Fea2010}
Friedman, J., Hastie, T., and Tibshirani, R. (2010).
\newblock Regularization paths for generalized linear models via coordinate
  descent.
\newblock {\em Journal of {S}tatistical {S}oftware}, 33(1):1.

\bibitem[Fu, 1998]{Fu1998}
Fu, W.~J. (1998).
\newblock Penalized regressions: the bridge versus the lasso.
\newblock {\em Journal of Computational and Graphical Statistics},
  7(3):397--416.

\bibitem[Jacob et~al., 2009]{Jea2009}
Jacob, L., Obozinski, G., and Vert, J.-P. (2009).
\newblock Group lasso with overlap and graph lasso.
\newblock In {\em Proceedings of the 26th {A}nnual {I}nternational {C}onference
  on {M}achine {L}earning}, pages 433--440.

\bibitem[Lange, 2010]{Lan2010}
Lange, K. (2010).
\newblock {\em Numerical Analysis for Statisticians}.
\newblock Springer Science \& Business Media.

\bibitem[Lee and Breheny, 2015]{LB2015}
Lee, S. and Breheny, P. (2015).
\newblock Strong rules for nonconvex penalties and their implications for
  efficient algorithms in high-dimensional regression.
\newblock {\em Journal of Computational and Graphical Statistics},
  24(4):1074--1091.

\bibitem[Lekivetz and Jones, 2015]{LJ2015}
Lekivetz, R. and Jones, B. (2015).
\newblock Fast flexible space-filling designs for nonrectangular regions.
\newblock {\em Quality and Reliability Engineering International},
  31(5):829--837.

\bibitem[Mak and Joseph, 2017]{MJ2016}
Mak, S. and Joseph, V.~R. (2017).
\newblock Minimax and minimax projection designs using clustering.
\newblock {\em Journal of Computational and Graphical Statistics}.
\newblock To appear.

\bibitem[Mazumder et~al., 2011]{Mea2011}
Mazumder, R., Friedman, J.~H., and Hastie, T. (2011).
\newblock Sparse{N}et: Coordinate descent with nonconvex penalties.
\newblock {\em Journal of the American Statistical Association},
  106(495):1125--1138.

\bibitem[Mazumder et~al., 2012]{Mea2012}
Mazumder, R., Hastie, T., and Friedman, J. (2012).
\newblock \textsc{sparsenet}: Fit sparse linear regression models via nonconvex
  optimization.
\newblock {\em R package version 1}.

\bibitem[Meier et~al., 2008]{Mea2008}
Meier, L., Van De~Geer, S., and B{\"u}hlmann, P. (2008).
\newblock The group lasso for logistic regression.
\newblock {\em Journal of the Royal Statistical Society: Series B},
  70(1):53--71.

\bibitem[Montgomery, 2008]{Mon2008}
Montgomery, D.~C. (2008).
\newblock {\em Design and {A}nalysis of {E}xperiments}.
\newblock John Wiley \& Sons.

\bibitem[Rosenbloom et~al., 1999]{Rea1999}
Rosenbloom, A.~L., Joe, J.~R., Young, R.~S., and Winter, W.~E. (1999).
\newblock Emerging epidemic of type 2 diabetes in youth.
\newblock {\em Diabetes {C}are}, 22(2):345--354.

\bibitem[Simon et~al., 2013]{Sea2013}
Simon, N., Friedman, J., Hastie, T., and Tibshirani, R. (2013).
\newblock A sparse-group lasso.
\newblock {\em Journal of Computational and Graphical Statistics},
  22(2):231--245.

\bibitem[Stuart and Ord, 1994]{SO1994}
Stuart, A. and Ord, J. (1994).
\newblock {\em Kendall's Advanced Theory of Statistics, Volume 1: Distribution
  Theory}.
\newblock Arnold London.

\bibitem[Su and Wu, 2017]{SW2016}
Su, H. and Wu, C. F.~J. (2017).
\newblock {CME} analysis: a new method for unraveling aliased effects in
  two-level fractional factorial experiments.
\newblock {\em Journal of Quality Technology}, 49(1):1--10.

\bibitem[Tibshirani, 1996]{Tib1996}
Tibshirani, R. (1996).
\newblock Regression shrinkage and selection via the lasso.
\newblock {\em Journal of the Royal Statistical Society: Series B},
  58(1):267--288.

\bibitem[Tibshirani, 1997]{Tib1997}
Tibshirani, R. (1997).
\newblock The lasso method for variable selection in the {C}ox model.
\newblock {\em Statistics in {M}edicine}, 16(4):385--395.

\bibitem[Tibshirani et~al., 2012]{Tea2012}
Tibshirani, R., Bien, J., Friedman, J., Hastie, T., Simon, N., Taylor, J., and
  Tibshirani, R.~J. (2012).
\newblock Strong rules for discarding predictors in lasso-type problems.
\newblock {\em Journal of the Royal Statistical Society: Series B},
  74(2):245--266.

\bibitem[Weber et~al., 2001]{Wea2001}
Weber, K., Eisman, R., Higgins, S., Morey, L., Patty, A., Tausek, M., and Zeng,
  Z.-B. (2001).
\newblock An analysis of polygenes affecting wing shape on chromosome 2 in
  {D}rosophila {M}elanogaster.
\newblock {\em Genetics}, 159(3):1045--1057.

\bibitem[Wu, 2015]{Wu2015}
Wu, C. F.~J. (2015).
\newblock Post-{F}isherian experimentation: from physical to virtual.
\newblock {\em Journal of the American Statistical Association},
  110(510):612--620.

\bibitem[Wu and Hamada, 2009]{WH2011}
Wu, C. F.~J. and Hamada, M.~S. (2009).
\newblock {\em Experiments: {P}lanning, {A}nalysis, and {O}ptimization}.
\newblock John Wiley \& Sons.

\bibitem[Wu and Lange, 2008]{WL2008}
Wu, T.~T. and Lange, K. (2008).
\newblock Coordinate descent algorithms for lasso penalized regression.
\newblock {\em The Annals of Applied Statistics}, 2(1):224--244.

\bibitem[Yuan and Lin, 2006]{YL2006}
Yuan, M. and Lin, Y. (2006).
\newblock Model selection and estimation in regression with grouped variables.
\newblock {\em Journal of the Royal Statistical Society: Series B},
  68(1):49--67.

\bibitem[Zhang, 2010]{Zha2010}
Zhang, C.-H. (2010).
\newblock Nearly unbiased variable selection under minimax concave penalty.
\newblock {\em The Annals of Statistics}, 38(2):894--942.

\bibitem[Zhao and Yu, 2006]{ZB2006}
Zhao, P. and Yu, B. (2006).
\newblock On model selection consistency of lasso.
\newblock {\em The Journal of Machine Learning Research}, 7:2541--2563.

\bibitem[Zou and Hastie, 2005]{ZH2005}
Zou, H. and Hastie, T. (2005).
\newblock Regularization and variable selection via the elastic net.
\newblock {\em Journal of the Royal Statistical Society: Series B},
  67(2):301--320.

\end{thebibliography}

\pagebreak
\setcounter{page}{1}

\begin{appendices}
\section{Proof of Theorem \ref{thm:corr}}
The proof of this requires a simple lemma on normal orthant probabilities:
\begin{lemma}
\citep{SO1994} Let $(X_1, \cdots, X_p)$ follow the equicorrelated normal distribution, with $\mathbb{E}(X_j) = 0$, $\mathbb{E}(X_j^2) = 1$ and $\mathbb{E}(X_j X_k) = \rho$ for all $j \neq k$, and let $p_m= \mathbb{P}(X_1 >  0 , \cdots, X_m > 0)$. Then:
\[p_2 = \frac{\sin^{-1}\rho}{2 \pi} + \frac{1}{4} \quad \text{and} \quad  p_3 = \frac{3\sin^{-1}\rho}{4\pi} + \frac{1}{8}.\]
\label{lem:orth}
\end{lemma}
\vspace{-1.0cm}
For the main proof, note that each row of the latent matrix $\bm{Z}$ is i.i.d., so it suffices to fix $n=1$ and explore the correlation amongst the scalar ME quantities $\tilde{x}_{1,A}$ and CME quantities $\tilde{x}_{1,A|B+}$. We denote these as $\tilde{x}_A$ and $\tilde{x}_{A|B+}$ for brevity. Under the latent equicorrelated distribution $\mathcal{N}\{\bm{0},\rho \bm{J} + (1-\rho) \bm{I}\}$, it is easy to show that $\mathbb{E}[\tilde{x}_A] = 0$ and $\text{Var}[\tilde{x}_A] = 1$. Moreover, the CME $\tilde{x}_{A|B+}$ can be conditionally decomposed as $\tilde{x}_{A|B+} \stackrel{d}{=} R[2p_2]$ if $\tilde{x}_{B} = +1$, and 0 if $\tilde{x}_{B} = -1$, where $R[q]$ is the Rademacher random variable taking on +1 w.p. $q \in [0,1]$ and -1 otherwise. From this, we get:
\begin{align*}
\mu_c &\equiv \mathbb{E}[ \tilde{x}_{A|B+}] = \mathbb{E}[ \mathbb{E}[ \tilde{x}_{A|B+} | \tilde{x}_B]] = \frac{1}{2}(4p_2 - 1), \\
\sigma_c^2 & \equiv \text{Var}[ \tilde{x}_{A|B+}]= \text{Var}[ \mathbb{E}[\tilde{x}_{A|B+} | \tilde{x}_B]] + \mathbb{E}[ \text{Var}[\tilde{x}_{A|B+} | \tilde{x}_B]] = \frac{1}{2} - \left( \frac{\sin^{-1}\rho}{\pi} \right)^2.
\end{align*}

Consider the correlation between the MEs $\tilde{x}_A$ and $\tilde{x}_B$. Note that $\tilde{x}_A\tilde{x}_B$ equals +1 when $\tilde{x}_A$ and $\tilde{x}_B$ have the same sign, and equals -1 otherwise. Letting $\mathbb{P}(++)$ be the probability of $(\tilde{x}_A,\tilde{x}_B)=(+1,+1)$ (with similar notation for $+-$, $-+$ and $--$), Lemma \ref{lem:orth} then gives:
\[\text{Corr}(\tilde{x}_A,\tilde{x}_B)=[\mathbb{P}(++) + \mathbb{P}(++)] - [\mathbb{P}(+-) + \mathbb{P}(-+) ] = 2p_2 - 2[1/2 - p_2] = \frac{2\sin^{-1}\rho}{\pi}.\]

Next, consider the two sibling CMEs $\tilde{x}_{A|B+}$ and $\tilde{x}_{A|C+}$. Note that $\tilde{x}_{A|B+}\tilde{x}_{A|C+}$ equals +1 when both $\tilde{x}_B = +1$ and $\tilde{x}_C = +1$, and equals 0 otherwise. It follows that:
\[\text{Corr}(\tilde{x}_{A|B+},\tilde{x}_{A|C+}) = \frac{1}{\sigma_c^2}[\mathbb{P}(++) - \mu_c^2] = \frac{1}{\sigma_c^2}[p_2 - \mu_c^2] = \frac{1}{\sigma^{2}_c}\left\{ - \left(\frac{\sin^{-1}\rho}{\pi}\right)^2 + \frac{\sin^{-1}\rho}{2\pi} + \frac{1}{4} \right\}.\]
\noindent The correlation for parent-child pairs can be proved in an analogous way.

Consider now the two cousin CMEs $\tilde{x}_{B|A+}$ and $\tilde{x}_{C|A+}$. Note that $\tilde{x}_{B|A+}\tilde{x}_{C|A+}$ equals +1 when $\tilde{x}_A=+1$ and $\tilde{x}_B=\tilde{x}_C$, $\tilde{x}_{B|A+}\tilde{x}_{C|A+}$ equals -1 when $\tilde{x}_A=+1$ and $\tilde{x}_B\neq\tilde{x}_C$, and equals 0 otherwise. We then have:
\begin{align*}
\text{Corr}(\tilde{x}_{B|A+},\tilde{x}_{C|A+}) &= \frac{1}{\sigma_c^2}\left[ \left\{ \mathbb{P}(+++)+\mathbb{P}(+--)\right\} - \left\{ \mathbb{P}(++-)+\mathbb{P}(++-)\right\} - \mu^2_c\right]\\
&= \frac{1}{\sigma_c^2}\left[ \left\{ \mathbb{P}(+++)+(\mathbb{P}(--) - \mathbb{P}(---))\right\} - 2\left\{ \mathbb{P}(++)-\mathbb{P}(+++)\right\} - \mu^2_c\right]\\
&= \frac{1}{\sigma_c^2}[2p_3 - p_2 - \mu^2_c] = \frac{1}{\sigma^{2}_c} \left\{ - \left(\frac{\sin^{-1}\rho}{\pi}\right)^2 + \frac{\sin^{-1}\rho}{\pi} \right\}.
\end{align*}

\section{Proof of Theorem \ref{thm:inc}}
Let $\bm{X} \in \mathbb{R}^{n \times p'}$ be the normalized model matrix consisting of all main effects and CMEs, where $p' = p + 4 {p \choose 2}$. By the strong law of large numbers, the sample covariance matrix $\bm{C}_n = {\bm{X}}^T {\bm{X}}/n$ converges elementwise to some matrix $\bm{C} \in \mathbb{R}^{p' \times p'}$ with unit diagonal entries and off-diagonal entries given in Theorem \ref{thm:corr}. Consider the following block partition of $\bm{C} = \begin{pmatrix}
\bm{C}_{11} & \bm{C}_{12}\\
\bm{C}_{21} & \bm{C}_{22}
\end{pmatrix}$,
where $\bm{C}_{11}$ is the block for the active set $\mathcal{A}$, and $\bm{C}_{22}$ the block for the remaining variables. \cite{ZB2006} proved that the LASSO is sign-selection consistent only when the (weak) \textit{irrepresentability condition} holds: $\forall \boldsymbol{\zeta} \in \{-1,+1\}^{p'}, \; |\bm{C}_{21}\bm{C}_{11}^{-1}\boldsymbol{\zeta}| < \bm{1}$ (this is a slight simplification of the original condition under the current i.i.d. setting). Hence, sign-selection inconsistency can be proven if $\exists \boldsymbol{\zeta} \in \{-1, +1\}^{p'}$ and an inactive effect $j$ satisfying:
\begin{equation}
|\bm{C}_{21,j} \bm{C}_{11}^{-1}\boldsymbol{\zeta}| \geq 1, \quad \text{where} \quad \bm{C}_{21,j} \text{ is the row corresponding to effect $j$.}
\label{eq:irrep}
\end{equation}

Consider first a model with only $q \geq 3$ active siblings of the form $A|B+$, $A|C-$, ..., $A|R-$. Using the same principles as in Theorem \ref{thm:corr}, $\bm{C}_{11}$ can be shown to be a $q \times q$ matrix with unit diagonal, $[(1/2 - p_2)-\mu_c^2]/\sigma^2_c$ for off-diagonal entries in the first row and column, and $\psi_{sib}(\rho)$ for all other off-diagonal entries \footnote{$\psi_{me}(\rho)$, $\psi_{sib}(\rho)$, $\psi_{pc}(\rho)$ and $\psi_{cou}(\rho)$ are the pairwise correlations in Theorem \ref{thm:corr} for main effects, siblings, parent-child pairs and cousins, respectively. $\tilde{\psi}(\rho) = \sin^{-1}(\rho)/(\pi\sigma_c)$ is the pairwise correlation between a CME and its conditioned effect.}. Letting $A$ be the inactive effect, we have $\bm{C}_{21,A} = \psi_{pc}(\rho) \bm{1}_q^T$, and letting $\boldsymbol{\zeta} = \bm{1}_q$, it follows that $|\bm{C}_{21,A} \bm{C}_{11}^{-1} \boldsymbol{\zeta}| \geq 1$ for $\rho \geq 0$. By \eqref{eq:irrep}, part (a) is proven.

Next, consider a model with only $q=2$ active main effects, say, $A$ and $-B$. From Theorem \ref{thm:corr}, $\bm{C}_{11}$ is a $q \times q$ matrix with unit diagonal and $-\psi_{me}(\rho)$ on the off-diagonals. Let $A|B-$ be the inactive effect, so $\bm{C}_{21,A|B-} = (\psi_{pc}(\rho), \tilde{\psi}(\rho))$. Taking $\boldsymbol{\zeta} = (1,1)^T$, $|\bm{C}_{21,A|B-} \bm{C}_{11}^{-1} \boldsymbol{\zeta}| \geq 1$ for $\rho \geq 0.27$, thereby proving selection inconsistency.

Lastly, consider a model with only $q \geq 6$ active cousins of the form $B|A+$, $C|A-$, ..., $R|A-$. Using the same principles as in Theorem \ref{thm:corr}, $\bm{C}_{11}$ is a $q \times q$ matrix with unit diagonal, $-\mu_c^2/\sigma_c^2$ for the off-diagonal entries in the first row and column, and $\psi_{cou}(\rho)$ for all other off-diagonal entries. Let $B$ be the inactive effect with $\bm{C}_{21,B} = (\psi_{sib}(\rho),\tilde{\psi}(\rho)\bm{1}_{q-1})$. Taking $\boldsymbol{\zeta} = \bm{1}_q$, $|\bm{C}_{21,B} \bm{C}_{11}^{-1} \boldsymbol{\zeta}| \geq 1$ for $\rho \geq 0.29$, which proves inconsistency.

\section{Proof of Proposition \ref{prop:conv}}
As a note, since the objective $Q(\boldsymbol{\beta})$ is non-differentiable at $\boldsymbol{\beta} = \bm{0}$, what we mean by strict convexity here is that $\nabla^2_{\bm{u}} Q(\boldsymbol{\beta})$, the directional Hessian of $Q(\boldsymbol{\beta})$ in direction $\bm{u}$, is positive-definite for all $\boldsymbol{\beta}$ and all $\|\bm{u}\| = 1$. We follow a similar approach as Proposition 1 of \cite{Bre2015}. Note that $\nabla^2 \|\bm{y}-\bm{X}\boldsymbol{\beta}\|^2_2 = 2 \bm{X}^T\bm{X}$. Moreover, with $\eta'_{\lambda,\tau}(\theta) = \lambda \exp(-\theta\tau/\lambda)$ and $\eta''_{\lambda,\tau}(\theta) = -\tau \exp(-\theta\tau/\lambda)$, one can show that $\nabla_{\bm{u}}^2 P_s(\boldsymbol{\beta}) \geq -\tau(1) + \lambda (-1/(\lambda\gamma)) = -\tau-1/\gamma$ and similarly $\nabla_{\bm{u}}^2 P_c(\boldsymbol{\beta}) \geq -\tau-1/\gamma$, for all $\bm{u}$ and $\boldsymbol{\beta}$. Hence:
\[ \nabla^2_{\bm{u}} Q(\boldsymbol{\beta}) = \nabla^2_{\bm{u}} \left\{ \frac{1}{2n}\|\bm{y}-\bm{X}\boldsymbol{\beta}\|^2_2 + P_s(\boldsymbol{\beta}) + P_c(\boldsymbol{\beta}) \right\} \geq \frac{\lambda_{min}(\bm{X}^T \bm{X})}{n} - 2 \left( \tau + \frac{1}{\gamma}\right) \text{ for all $\bm{u}$ and $\boldsymbol{\beta}$,}\]
which is strictly positive when $\tau + 1/\gamma < \lambda_{min}(\bm{X}^T \bm{X})/(2n)$. The second part of the claim follows by replacing $\bm{X}$ with $\bm{x}_j$ in the argument above, and using the fact that $\|\bm{x}_j\|_2^2 = n$.\\

\section{Proof of Theorem \ref{thm:thresh} and Corollary \ref{cor:conv}}
The majorization claim $a)$ follows from a first-order Taylor expansion of the outer penalty: $\eta_{\lambda,\tau}(\|\boldsymbol{\beta}_g\|_{\lambda,\gamma}) \geq \eta_{\lambda,\tau}(\|\tilde{\boldsymbol{\beta}}_g\|_{\lambda,\gamma}) + \tilde{\Delta}_g \left\{ \|{\boldsymbol{\beta}}_g\|_{\lambda,\gamma} - \|\tilde{\boldsymbol{\beta}}_g\|_{\lambda,\gamma} \right\}$, where the inequality holds due to the concavity of $\eta$. See Lemma 1 in \cite{Bre2015} for details.

To derive the threshold function in $b)$, take the following optimization problem:
\begin{equation}
\hat{\beta}_j = \argmin_{\beta_j} \left\{ \frac{1}{2n}\| {\bm{r}} - \bm{x}_j \beta_j\|^2_2 + {\Delta}_1 g_{\lambda_1,\gamma}(\beta_j)  + {\Delta}_2 g_{\lambda_2,\gamma}(\beta_j) \right\}.
\label{eq:majsol}
\end{equation}
The KKT condition for \eqref{eq:majsol} is:
\begin{equation}
0 \in -\frac{1}{n} \bm{x}_j^T\bm{r} + \hat{\beta}_j + \Delta_1 \partial_{\lambda_1, \gamma} \hat{\beta}_j + \Delta_2 \partial_{\lambda_2, \gamma} \hat{\beta}_j, \quad \partial_{\lambda, \gamma} \beta_j = \begin{dcases}
\text{sgn}(\beta_j) \left( 1 - \frac{|\beta_j|}{\lambda \gamma} \right)_+ & \text{if } |\beta_j| > 0,\\
[-1,1] & \text{if } \beta_j = 0.
\end{dcases}
\label{eq:majsolkkt}
\end{equation}
Without loss of generality, assume $z \equiv \bm{x}_j^T \bm{r}/n > 0$. Consider the same four cases for $z$ as presented in \eqref{eq:thresh}:
\ben
\item $z \geq \lambda_{(1)} \gamma$: Suppose $\hat{\beta}_j = z$. Then the KKT condition \eqref{eq:majsolkkt} becomes $0 \in - z + \hat{\beta}_j$, which is satisfied. Since \eqref{eq:majsol} is strictly convex, $\hat{\beta}_j = z$ must be its unique solution.
\item $c_2 \leq z < \lambda_{(1)} \gamma$ (see \eqref{eq:thresh} for $c_2$): Suppose $\hat{\beta}_j = (z - \Delta_{(1)})/\left( 1 - \frac{\Delta_{(1)}}{\lambda_{(1)} \gamma} \right)$. Since $\lambda_{(2)} \gamma \leq \hat{\beta}_j < \lambda_{(1)}\gamma$, the KKT condition \eqref{eq:majsolkkt} becomes $0 \in - z + \hat{\beta}_j + \Delta_{(1)} \left( 1 - \frac{\hat{\beta}_j}{\lambda_{(1)}\gamma} \right)$, which is satisfied. Hence, $\hat{\beta}_j$ is the unique solution to \eqref{eq:majsol}.
\item $\Delta_{(1)} + \Delta_{(2)} \leq z < c_2$ (see \eqref{eq:thresh} for $c_3$): Suppose $\hat{\beta}_j = (z - \Delta_{(1)} - \Delta_{(2)})/\left( 1 - \frac{\Delta_{(1)}}{\lambda_{(1)} \gamma} - \frac{\Delta_{(2)}}{\lambda_{(2)} \gamma}\right)$. Since $0 < \hat{\beta}_j < \lambda_{(2)}\gamma$, the KKT condition \eqref{eq:majsolkkt} becomes $0 \in - z + \hat{\beta}_j + \Delta_{(1)} \left( 1 - \frac{\hat{\beta}_j}{\lambda_{(1)}\gamma} \right) + \Delta_{(2)} \left( 1 - \frac{\hat{\beta}_j}{\lambda_{(2)}\gamma} \right)$, which is satisfied. Hence, $\hat{\beta}_j$ is the unique solution to \eqref{eq:majsol}.
\item $0 \leq z < \Delta_{(1)} + \Delta_{(2)}$: Suppose $\hat{\beta}_j = 0$. The KKT condition then becomes $0 \in -z + (\Delta_{(1)} + \Delta_{(2)})[-1,1]$, which is satisfied, so $\hat{\beta}_j$ is the unique solution to \eqref{eq:majsol}.
\een

From this, Corollary \ref{cor:conv} can be proved in a similar way as Proposition 3 of \cite{Bre2015}.

\section{Proof of Proposition \ref{prop:maxl}}
Since $Q(\boldsymbol{\beta})$ is strictly convex, it must have at most one minimizer $\boldsymbol{\beta}$. By definition, $\boldsymbol{\beta}$ must satisfy the KKT condition:
\begin{equation}
0 \in -\frac{1}{n}\bm{x}_j^T(\bm{y} - \bm{X}\boldsymbol{\beta}) + \Delta_{\mathcal{S}}(\boldsymbol{\beta}) \partial_{\lambda_s, \gamma} \beta_j +  \Delta_{\mathcal{C}}(\boldsymbol{\beta}) \partial_{\lambda_c, \gamma} \beta_j, \quad j = 1, \cdots, p',
\label{eq:subgrad}
\end{equation}
where $\partial_{\lambda,\gamma} \beta_j$ is the subgradient defined in \eqref{eq:majsolkkt}, and $\Delta_{\mathcal{S}}(\boldsymbol{\beta})$ and $\Delta_{\mathcal{C}}(\boldsymbol{\beta})$ are the linearized slopes in \eqref{eq:couple} for the sibling and cousin groups of effect $j$. Setting $\beta = \bm{0}$, the right side of \eqref{eq:subgrad} becomes:
\[-\frac{1}{n}\bm{x}_j^T\bm{y} + \lambda_s [-1, 1] + \lambda_c [-1,1]  = -\frac{1}{n}\bm{x}_j^T\bm{y} + [-\lambda_s - \lambda_c, \lambda_s + \lambda_c],\]
which contains 0 when $\lambda_s + \lambda_c \geq |\bm{x}_j^T\bm{y}|/n$. Hence, when $\lambda_s + \lambda_c \geq \max_{j=1, \cdots, p'} |\bm{x}_j^T \bm{y}|/n$, one can invoke the strict convexity of $Q(\boldsymbol{\beta})$ to show that the trivial solution $\boldsymbol{\beta} =\bm{0}$ is indeed the unique minimizer.

\section{Algorithm statement for \texttt{cv.cmenet}}
\label{sec:cvcmenet}
\begin{algorithm}[!h]
\caption{\texttt{cv.cmenet}: A cross-validation algorithm for tuning \texttt{cmenet}}
\label{alg:cvcmenet}
\begin{algorithmic}[1]
\small
\Function{\texttt{cv.cmenet}}{$\bm{X},\bm{y},K$}
\stb Initialize grid of potential parameters $\displaystyle \max_{j=1,\cdots,p'} | \bm{x}_j^T \bm{y}|/n > \lambda_s^1 > \cdots > \lambda_s^L > 0$, $\displaystyle \max_{j=1,\cdots,p'} | \bm{x}_j^T \bm{y}|/n > \lambda_c^1 > \cdots > \lambda_c^M > 0$, $\gamma^1 < \cdots < \gamma^G$ and $\tau^1 < \cdots < \tau^T$ (satisfying $\tau + 1/\gamma < 1/2$).
\stb Obtain the tuned MC+ parameters $(\lambda^*,\gamma^*)$ using \texttt{cv.sparsenet} in the R package \textsc{sparsenet}, and set $\lambda_s^*, \lambda_c^* \leftarrow \lambda^*/2$ as an initial estimate.
\stb Randomly partition the data $\mathcal{D} = (\bm{X},y)$ into $K$ equal pieces $\{\mathcal{D}_1, \cdots, \mathcal{D}_K\}$.
\For{$k = 1, \cdots, K$} \Comment{$K$-fold CV for tuning $\gamma$ and $\tau$}
\For{$\gamma \in \{\gamma_1, \cdots, \gamma_G\}$} \Comment{For each $\gamma$...}
\stb $\boldsymbol{\beta}_{prev} \leftarrow \bm{0}_{p'}$ \Comment{Reset warm start solution}
\For{$\tau \in \{\tau_1, \cdots, \tau_T\}$} \Comment{For each $\tau$...}
\stb $\boldsymbol{\beta}_{\lambda_s^*, \lambda_c^*}(\gamma,\tau; k) \leftarrow \texttt{cmenet}(\bm{X}_{-k},\bm{y}_{-k},\lambda_s^*,\lambda_c^*,\gamma,\tau,\boldsymbol{\beta}_{prev})$ \Comment{Train w/o part $k$}
\stb $\boldsymbol{\beta}_{prev} \leftarrow \boldsymbol{\beta}_{\lambda_s^*, \lambda_c^*}(\gamma,\tau; k)$ \Comment{Update warm start solution}
\EndFor
\EndFor
\EndFor
\stb $(\gamma^*, \tau^*)\leftarrow \displaystyle \argmin_{\gamma,\tau}\sum_{k=1}^K \|\bm{y}_k - \bm{X}_k \boldsymbol{\beta}_{\lambda_s^*, \lambda_c^*}(\gamma,\tau;k)\|_2^2$ \Comment{Estimate optimal $\gamma$ and $\tau$}
\For{$k = 1, \cdots, K$} \Comment{$K$-fold CV for tuning $\lambda_s$ and $\lambda_c$}
\For{$\lambda_c \in \{\lambda_c^1, \cdots, \lambda_c^M\}$} \Comment{For each $\lambda_c$...}
\stb $\boldsymbol{\beta}_{prev} \leftarrow \bm{0}_{p'}$
\For{$\lambda_s \in \{\lambda_s^1, \cdots, \lambda_s^L\}$} \Comment{For each $\lambda_s$...}
\If{$\lambda_c + \lambda_s < \max_{j=1,\cdots,p'} | \bm{x}_j^T \bm{y}| /n$}
\stb Screen using the three strong rules in Section \ref{sec:screen}.
\stb $\boldsymbol{\beta}_{\lambda_s, \lambda_c}(\gamma^*,\tau^*; k) \leftarrow \texttt{cmenet}(\bm{X}_{-k},\bm{y}_{-k},\lambda_s,\lambda_c,\gamma^*,\tau^*,\boldsymbol{\beta}_{prev})$,\par
\hskip \algorithmicindent \hskip \algorithmicindent \hskip \algorithmicindent \hskip \algorithmicindent using only screened effects.
\stb Check KKT conditions on converged solution $\boldsymbol{\beta}_{\lambda_s, \lambda_c}(\gamma^*,\tau^*; k)$.
\stb $\boldsymbol{\beta}_{prev} \leftarrow \boldsymbol{\beta}_{\lambda_s, \lambda_c}(\gamma^*,\tau^*; k)$ 
\EndIf
\EndFor
\EndFor
\EndFor
\stb $(\lambda_s^*, \lambda_c^*)\leftarrow \displaystyle \argmin_{\lambda_s,\lambda_c}\sum_{k=1}^K \|\bm{y}_k - \bm{X}_k \boldsymbol{\beta}_{\lambda_s, \lambda_c}(\gamma^*,\tau^*;k)\|_2^2$ \Comment{Estimate optimal $\lambda_s$ and $\lambda_c$}
\stb $\hat{\boldsymbol{\beta}} \leftarrow \texttt{cmenet}(\bm{X},\bm{y},\lambda_s^*,\lambda_c^*,\gamma^*,\tau^*,\bm{0}_{p'})$ \Comment{Refit using optimal parameters}\par
\Return optimal coefficients $\hat{\boldsymbol{\beta}}$.
\EndFunction
\normalsize
\end{algorithmic}
\end{algorithm}

Some comments on the implementation of active set optimization within \texttt{cmenet}:
\bi
\item The active set of variables is initialized by performing the full coordinate descent cycle for 25 iterations, then choosing the variables whose coefficients are non-zero.
\item Repeat coordinate descent iterations over the active set until convergence.
\item Perform a full coordinate descent cycle over all $p'$ variables. If this cycle does not change the active set, \texttt{cmenet} is terminated; otherwise, the active set is updated, and the above steps repeated.
\ei


\section{Theoretical derivation of CME screening rules}
\label{sec:cmescreen}
Fix $\gamma$ and $\tau$, and suppose $\hat{\beta}_j(\lambda_s,\lambda_c) \in (0,\min\{\Delta_{(1)}+\Delta_{(2)}, \lambda_{(2)}\gamma\})$. For brevity, we denote $\hat{\beta}_j(\lambda_s,\lambda_c)$ as $\hat{\beta}_j$ from here on. Using equation \eqref{eq:thresh}, we know that $\hat{\beta}_j$ takes the form:
\begin{align}
\small
\begin{split}
\hat{\beta}_j &= \textup{sgn}(z_j) \left(|z_j| - \Delta_{(1)} - \Delta_{(2)} \right)_+ / \left(1 - \frac{\Delta_{(1)}}{\lambda_{(1)} \gamma} - \frac{\Delta_{(2)}}{\lambda_{(2)} \gamma} \right)\\
& = \textup{sgn}(z_j) \left(|z_j| - \Delta_{S} - \Delta_{C} \right)_+ / \left(1 - \frac{\Delta_{S}}{\lambda_{S} \gamma} - \frac{\Delta_{C}}{\lambda_{C} \gamma} \right),
\label{eq:bj}
\end{split}
\normalsize
\end{align}
where $z_j = \bm{x}_j^T\bm{r}_{-j}/n$ (see Theorem \ref{thm:thresh}), and $\Delta_{\mathcal{S}}$ and $\Delta_{\mathcal{C}}$ are the linearized slopes for the current penalty setting $(\lambda_s,\lambda_c)$. Plugging this expression into \eqref{eq:subgrad}, the KKT condition for $\hat{\beta}_j$ can be simplified to:
\small
\begin{align}
\begin{split}
0 = -c_j(\lambda_s, \lambda_c) + \text{sgn}(\hat{\beta}_j) \Delta_{\mathcal{S}} \left\{1 - \frac{(|z_j|-\Delta_{\mathcal{S}}-\Delta_{\mathcal{C}})_+}{ \lambda_s \left(\gamma - \frac{\Delta_{\mathcal{S}}}{\lambda_s} - \frac{\Delta_{\mathcal{C}}}{\lambda_c} \right)}\right\} + \text{sgn}(\hat{\beta}_j) \Delta_{\mathcal{C}} \left\{1 - \frac{(|z_j|-\Delta_{\mathcal{S}}-\Delta_{\mathcal{C}})_+}{\lambda_c \left( \gamma - \frac{\Delta_{\mathcal{S}}}{\lambda_s} - \frac{\Delta_{\mathcal{C}}}{\lambda_c}\right)}\right\}\\
\Leftrightarrow \; c_j(\lambda_s, \lambda_c) = \text{sgn}(\hat{\beta}_j) \Delta_{\mathcal{S}} \left\{1 - \frac{(|z_j|-\Delta_{\mathcal{S}}-\Delta_{\mathcal{C}})_+}{ \lambda_s \left(\gamma - \frac{\Delta_{\mathcal{S}}}{\lambda_s} - \frac{\Delta_{\mathcal{C}}}{\lambda_c} \right)}\right\} + \text{sgn}(\hat{\beta}_j) \Delta_{\mathcal{C}} \left\{1 - \frac{(|z_j|-\Delta_{\mathcal{S}}-\Delta_{\mathcal{C}})_+}{\lambda_c \left( \gamma - \frac{\Delta_{\mathcal{S}}}{\lambda_s} - \frac{\Delta_{\mathcal{C}}}{\lambda_c}\right)}\right\}.
\end{split}
\label{eq:kktscreen}
\end{align}
\normalsize
\vspace{-1.0cm}

Suppose no effects are active in either the sibling group $\mathcal{S}$ or the cousin group $\mathcal{C}$, in which case $\Delta_{\mathcal{S}} = \lambda_s$ and $\Delta_{\mathcal{C}} = \lambda_c$. The KKT condition in \eqref{eq:kktscreen} can then be rewritten as:
\begin{equation}
c_j(\lambda_s, \lambda_c) = \text{sgn}(\hat{\beta}_j) \left\{\lambda_s - \frac{(|z_j|-\lambda_s-\lambda_c)_+}{\gamma - 2}\right\} + \text{sgn}(\hat{\beta}_j) \left\{\lambda_c - \frac{(|z_j|-\lambda_s - \lambda_c)_+}{\gamma -2}\right\}.
\end{equation}
Taking the derivative with respect to $\lambda_s$ (and assuming $z_j$ is approximately constant in $\lambda_s$, following \citealp{LB2015}), we get:
\begin{equation}
\Big| \frac{\partial}{\partial \lambda_s} c_j(\lambda_s,\lambda_c) \Big| \lesssim 1 + \frac{1}{\gamma-2} + \frac{1}{\gamma-2} = \frac{\gamma}{\gamma-2}.
\label{eq:sc1up}
\end{equation}
A similar argument shows that this approximate upper bound also holds for $| ({\partial}/{\partial \lambda_c}) \; c_j(\lambda_s,\lambda_c)|$.

Now, suppose no effects are active in the sibling group $\mathcal{S}$ (but some in the cousin group $\mathcal{C}$), in which case $\Delta_{\mathcal{S}} = \lambda_s$. The KKT condition in \eqref{eq:kktscreen} can then be rewritten as:
\vspace{-0.25cm}
\begin{equation}
c_j(\lambda_s, \lambda_c) = \text{sgn}(\hat{\beta}_j) \left\{\lambda_s - \frac{(|z_j|-\lambda_s-\Delta_{\mathcal{C}})_+}{\gamma - 1 - \frac{\Delta_{\mathcal{C}}}{\lambda_c}}\right\} + \text{sgn}(\hat{\beta}_j) \Delta_{\mathcal{C}} \left\{1 - \frac{(|z_j|-\lambda_s - \Delta_{\mathcal{C}})_+}{\lambda_c \left( \gamma - 1 - \frac{\Delta_{\mathcal{C}}}{\lambda_c} \right)}\right\}.
\end{equation}
\vspace{-0.25cm}
\noindent Taking the derivative on $\lambda_s$ (and assuming $z_j$ is approximately constant in $\lambda_s$), we get:
\begin{equation}
\Big| \frac{\partial}{\partial \lambda_s} c_j(\lambda_s,\lambda_c) \Big| \lesssim 1 + \frac{1}{\gamma-1-\frac{\Delta_{\mathcal{C}}}{\lambda_c}} + \frac{\frac{\Delta_{\mathcal{C}}}{\lambda_c}}{\gamma-1-\frac{\Delta_{\mathcal{C}}}{\lambda_c}} = \frac{\gamma}{\gamma-1-\frac{\Delta_{\mathcal{C}}}{\lambda_c}}.
\label{eq:sc2up}
\end{equation}
Finally, suppose there are no active effects in the cousin group $\mathcal{C}$ (but some in sibling group $\mathcal{S}$). One can do a similar approximation and show that:
\begin{equation}
\Big| \frac{\partial}{\partial \lambda_c} c_j(\lambda_s,\lambda_c) \Big| \lesssim 1 + \frac{1}{\gamma-\frac{\Delta_{\mathcal{S}}}{\lambda_s}-1} + \frac{\frac{\Delta_{\mathcal{S}}}{\lambda_s}}{\gamma-\frac{\Delta_{\mathcal{S}}}{\lambda_s}-1} = \frac{\gamma}{\gamma-\frac{\Delta_{\mathcal{S}}}{\lambda_s}-1}.
\label{eq:sc3up}
\end{equation}

These upper bounds on the absolute derivatives of $c_j(\lambda_s,\lambda_c)$, along with the proposed strong rules in Section \ref{sec:screen}, can then be used to demonstrate the inactivity of effect $j$ at penalty setting $(\lambda_s^l,\lambda_c^m)$:
\ben
\item Consider the first part of the first strong rule, which applies when no active effects are in $\mathcal{S}$ and $\mathcal{C}$ for setting $(\lambda_s^{l-1},\lambda_c^m)$. This rule discards effect $j$ at setting $(\lambda_s^l,\lambda_c^m)$ if:
\[|c_j(\lambda_s^{l-1},\lambda_c^{m})| < \lambda_s^{l} + \lambda_c^m + \frac{\gamma}{\gamma-2} (\lambda_s^l - \lambda_s^{l-1}).\]
This can be justified as follows. Using the approximate upper bound in \eqref{eq:sc1up}, the inner-product of effect $j$ at setting $(\lambda_s^l,\lambda_c^m)$ can be approximately upper bounded as:
\begin{align*}
|c_j(\lambda_s^l,\lambda_c^m)| &\leq |c_j(\lambda_s^l,\lambda_c^m) - c_j(\lambda_s^{l-1},\lambda_c^m ) | + |c_j(\lambda_s^{l-1},\lambda_c^{m} )|\\
& \approx \Big| \frac{\partial}{\partial \lambda_s} c_j(\lambda_s^{l-1},\lambda_c^{m}) \Big| (\lambda_s^{l-1} - \lambda_s^l) + |c_j(\lambda_s^{l-1},\lambda_c^{m} )|\\
& < \frac{\gamma}{\gamma-2}(\lambda_s^{l-1} - \lambda_s^{l}) + \left[ \lambda_s^{l} + \lambda_c^{m} + \frac{\gamma}{\gamma-2} (\lambda_s^l - \lambda_s^{l-1}) \right]\\
& = \lambda_s^l + \lambda_c^m.
\end{align*}
Assuming effect $j$ is the first variable to potentially be selected in $\mathcal{S}$ or $\mathcal{C}$ at current setting $(\lambda_s^l,\lambda_c^m)$, the KKT conditions in \eqref{eq:subgrad} suggest that effect $j$ is inactive, which justifies the screening rule. A similar argument can be used to derive the second part of this rule.
\item Consider next the second strong rule, which applies when no active effects are in $\mathcal{S}$ for setting $(\lambda_s^{l-1}, \lambda_c^m)$. This rule discards effect $j$ at setting $(\lambda_s^l,\lambda_c^m)$ if:
\[|c_j(\lambda_s^{l-1},\lambda_c^{m})| < \lambda_s^{l} + \Delta_{\mathcal{C}}' + \frac{\gamma}{\gamma-(\Delta_{\mathcal{C}}'/\lambda_c^m+1)} (\lambda_s^l - \lambda_s^{l-1}).\]
This can be justified as follows. Using the approximate upper bound in \eqref{eq:sc2up}, the inner-product of effect $j$ at setting $(\lambda_s^l,\lambda_c^m)$ can be approximately upper bounded as:
\begin{align*}
|c_j(\lambda_s^l,\lambda_c^m)| &\leq |c_j(\lambda_s^l,\lambda_c^m) - c_j(\lambda_s^{l-1},\lambda_c^m ) | + |c_j(\lambda_s^{l-1},\lambda_c^{m} )|\\
& \approx \Big| \frac{\partial}{\partial \lambda_s} c_j(\lambda_s^{l-1},\lambda_c^{m}) \Big| (\lambda_s^{l-1} - \lambda_s^l) + |c_j(\lambda_s^{l-1},\lambda_c^{m} )|\\
& < \frac{\gamma}{\gamma-(\Delta_{\mathcal{C}}'/\lambda_c^m+1)}(\lambda_s^{l-1} - \lambda_s^{l}) + \left[ \lambda_s^{l} + \Delta_{\mathcal{C}}' + \frac{\gamma}{\gamma-(\Delta_{\mathcal{C}}'/\lambda_c^m +1)} (\lambda_s^l - \lambda_s^{l-1}) \right]\\
& = \lambda_s^l + \Delta_{\mathcal{C}}'.
\end{align*}
Assuming:
\bi
\item Effect $j$ is the first variable to potentially be selected in $\mathcal{S}$ at current setting $(\lambda_s^l,\lambda_c^m)$,
\item The linearized slope $\Delta_{\mathcal{C}}'$ at previous setting $(\lambda_s^{l-1},\lambda_c^m)$ is approximately the linearized slope $\Delta_{\mathcal{C}}$ at current setting $(\lambda_s^l,\lambda_c^m)$,
\ei
the KKT conditions in \eqref{eq:subgrad} suggest that effect $j$ is inactive, which justifies the screening rule. 
\item The third strong rule can be justified in a similar manner to the above two rules.
\een

\end{appendices}

\end{document}